\newif\ifanonymous

\anonymousfalse

\newif\ifarxiv
\arxivtrue

\ifanonymous
\documentclass[manuscript,sigconf,screen,review,anonymous]{acmart}
\else
\documentclass[sigconf,screen]{acmart}%\settopmatter{printfolios=true,printccs=true,printacmref=false}
\fi
\AtBeginDocument{%
  }

\usepackage{graphicx}
\graphicspath{{./fig/}}
\DeclareGraphicsExtensions{.pdf,.eps}
\usepackage{picinpar} % for using wrapfigure in theorem-like environments, e.g. example.
\usepackage{wrapfig}
\usepackage{subcaption}
\usepackage{adjustbox}
\usepackage{url}
\usepackage{amsmath,stmaryrd,amsthm}
\usepackage{thm-restate}
\usepackage{nccmath} %for reducing font-size in the align environment, load before mathtools
\usepackage{nicefrac}

\usepackage{fancyvrb}
\usepackage{xspace}
\usepackage{threeparttable}

\usepackage{mathtools}
\usepackage[long]{optidef}
\usepackage{csquotes}
\usepackage[inline]{enumitem}
\usepackage{etoolbox}

\usepackage[ruled,vlined,linesnumbered]{algorithm2e}
\SetKwRepeat{Do}{do}{while}
\SetKwComment{tcp}{$\rhd\;$}{}%
\SetAlCapFnt{\small}             % font for caption text
\SetAlCapNameFnt{\small\bfseries}% font for "Algorithm" label
\SetAlFnt{\footnotesize}         % font for the algorithm body (pseudocode)

\makeatletter
\patchcmd{\@algocf@start}% <cmd>
{-1.5em}% <search>
{0pt}% <replace>
{}{}% <success><failure>
\makeatother

\usepackage{pifont}% http://ctan.org/pkg/pifont

\usepackage[font=small,labelfont=bf]{caption}

\usepackage{comment}
\usepackage{algpseudocode}
\usepackage{float}
\usepackage{tcolorbox}
\usepackage{booktabs}
\usepackage{marvosym}
\usepackage{fontawesome}

\usepackage{makecell}
\usepackage{tikz}
\usetikzlibrary{calc,arrows,shapes,snakes,automata,backgrounds,petri,positioning}
\tikzset{mynode/.style={draw,solid,circle,inner sep=1pt}}

\usepackage{nicematrix}
\NiceMatrixOptions{cell-space-top-limit = 1pt, cell-space-bottom-limit = 1pt}

\AtBeginEnvironment{pmatrix}{\setlength{\arraycolsep}{4pt}}

\usepackage{array}
\usepackage{multirow}
\usepackage{bm}

\usepackage{scalerel} %for scaling sub-/supscripts in math

\usepackage[capitalize]{cleveref}
\crefname{section}{Sect.}{Sect.}
\Crefname{section}{Sect.}{Sect.}

\makeatletter
\AddToHook{cmd/appendix/before}{\def\cref@section@alias{appendix}\def\cref@subsection@alias{appendix}}
\makeatother

\crefname{appendix}{Appx.}{Appx.}
\Crefname{appendix}{Appx.}{Appx.}
\crefname{equation}{}{}
\crefname{enumi}{}{}
\crefname{enumi}{Step}{Steps}
\Crefname{figure}{Fig.}{Figs.}
\crefname{prob}{problem}{problems}
\Crefname{prob}{Problem}{Problems}
\creflabelformat{prob}{#2\textup{(#1)}#3}

\crefname{cha}{challenge}{challenges}
\Crefname{cha}{Challenge}{Challenges}

\crefdefaultlabelformat{#2\textup{#1}#3}

\theoremstyle{plain}
\newtheorem*{thm*}{Theorem}
\newtheorem*{exmp*}{Example}
\newtheorem*{counterexmp*}{Counterexample}

\newtheoremstyle{ourstyle}{}{}{\itshape}{}{\bfseries}{.}{ }{\thmname{#1}\thmnumber{ #2}\thmnote{ (#3)}}
\theoremstyle{ourstyle}

\newtheorem{theorem}{Theorem}
\newtheorem{definition}[theorem]{Definition}
\newtheorem{lemma}[theorem]{Lemma}

\newtheorem{property}[theorem]{Property}

\newtheoremstyle{exmpstyle}{}{}{}{}{\bfseries}{.}{ }{\thmname{#1}\thmnumber{ #2}\thmnote{ (#3)}}
\theoremstyle{exmpstyle}

\newtheorem{example}[theorem]{Example}

\newtheoremstyle{rmkstyle}{}{}{}{}{\bfseries}{.}{ }{\thmname{#1}\thmnote{ (#3)}}
\theoremstyle{rmkstyle}

\newtheorem{remark}{Remark}
\input{def}
\copyrightyear{2026}
\acmYear{2026}
\setcopyright{cc}
\setcctype{by}
\acmConference[HSCC '26]{29th ACM International Conference on Hybrid Systems: Computation and Control}{May 11--14, 2026}{Saint Malo, France}
\acmBooktitle{29th ACM International Conference on Hybrid Systems: Computation and Control (HSCC '26), May 11--14, 2026, Saint Malo, France}
\acmDOI{10.1145/3801146.3805664}
\acmISBN{979-8-4007-2566-1/2026/05}

\begin{document}

\title{Derivative-Agnostic Inference of Nonlinear Hybrid Systems}

\settopmatter{authorsperrow=4}

\author{Hengzhi Yu}
\authornote{Both authors contributed equally to this research.}
\email{hengzhiyu@zju.edu.cn}
\orcid{0009-0003-7034-890X}
\affiliation{%
  \institution{Zhejiang University}
  \city{Hangzhou}
  \country{China}
}

\author{Bohan Ma}
\authornotemark[1]
\email{bohanma@zju.edu.cn}
\orcid{0009-0007-7036-997X}
\affiliation{%
  \institution{Zhejiang University}
  \city{Hangzhou}
  \country{China}
}

\author{Mingshuai Chen}
\authornote{The corresponding author.}
\email{m.chen@zju.edu.cn}
\orcid{0000-0001-9663-7441}
\affiliation{%
  \institution{Zhejiang University}
  \city{Hangzhou}
  \country{China}
}

\author{Huangying Dong}
\email{hydong@zju.edu.cn}
\orcid{0009-0002-0267-4466}
\affiliation{%
  \institution{Zhejiang University}
  \city{Hangzhou}
  \country{China}
}

\author{Jie An}
%\email{anjie@iscas.ac.cn}
\orcid{0000-0001-9260-9697}
\affiliation{%
  \institution{Institute of Software, CAS}
  \city{Beijing}
  \country{China}
}

\author{Bin Gu}
%\email{gubin@ios.ac.cn}
\orcid{0000-0001-7218-4839}
\affiliation{%
  \institution{Beijing Institute of Control Engineering, Beijing, China}
  \city{}
  \country{}
}

\author{Naijun Zhan}
%\email{njzhan@pku.edu.cn}
\orcid{0000-0003-3298-3817}
\affiliation{%
  \institution{Peking University}
  \city{Beijing}
  \country{China}
}

\author{Jianwei Yin}
%\email{zjuyjw@zju.edu.cn}
\orcid{0000-0003-4703-7348}
\affiliation{%
  \institution{Zhejiang University}
  \city{Hangzhou}
  \country{China}
}
\renewcommand{\shortauthors}{Yu et al.}

% !TEX root = main.tex

\begin{abstract}
This paper addresses the problem of inferring hybrid automata from input-output traces of hybrid systems exhibiting discrete mode switches between continuously evolving dynamics. Existing approaches primarily rely on derivative-based strategies in which (i) mode switches are detected by drastic variations in derivatives and (ii) trace segments are clustered based on signal similarity -- both requiring user-supplied thresholds. We present a derivative-agnostic approach, named {\ourmethod}, for inferring nonlinear hybrid systems whose dynamics are captured by nonlinear autoregressive exogenous (NARX) models. {\ourmethod} employs NARX models as a unified, threshold-free representation for both mode switching and segment clustering. We show that {\ourmethod} suffices to learn models that closely approximate a general class of hybrid systems featuring high-order nonlinear dynamics with exogenous inputs, nonlinear guard conditions, linear resets, and noise. Experimental results on a collection of benchmarks demonstrate that our approach effectively and efficiently infers nontrivial hybrid automata with high-order dynamics, yielding significantly more accurate approximations than state-of-the-art techniques, while achieving robustness comparable to approaches dedicated to fitting noisy data.
\end{abstract}


% \keywords{hybrid system identification, model/automata learning, nonlinear dynamics}

\setlength{\floatsep}{1\baselineskip}
\setlength{\textfloatsep}{1\baselineskip}
\setlength{\intextsep}{1\baselineskip}

\maketitle

% !TEX root = main.tex

\section{Introduction}

Hybrid systems are a commonly adopted mathematical model for cyber-physical systems (CPS). It describes the interaction between embedded computers and their continuously evolving physical environment or plant via sensors and actuators. Since \emph{model-based design} \cite{DBLP:conf/iwcmc/JensenCL11,DBLP:journals/pieee/KarsaiSLB03}
%\cmscomment{The latter reference can be removed if needed.} 
emerges as a predominant approach to the construction of CPS -- where a prerequisite is to have a model of the system under construction -- identifying the hybrid system model underneath a potentially black-box system through external observations has become an essential problem in the design of CPS. The general scope of this problem, known as \emph{system identification} \cite{DBLP:reference/sc/Ljung15} or \emph{model learning} \cite{Vaandrager17}, has undergone a surge of interest due to prominent applications in, e.g., system analysis~\cite{WatanabeFHLP0023,BartocciDGMNQ20,Shea-BlymyerA21,Fiterau-Brostean16}, model checking~\cite{Fiterau-Brostean17,PeledVY02,ShijuboWS21}, falsification~\cite{Waga20,BakBHKLMR24}, and monitoring~\cite{GhoshA24,JungesST24}.

Hybrid automata (HAs) \cite{DBLP:conf/rex/MalerMP91,DBLP:conf/hybrid/AlurCHH92} are a widely used formalism for modeling the discrete-continuous behaviors of hybrid systems. They extend finite-state machines by associating each discrete location (mode) with a vector of continuously evolving dynamics -- typically governed by %a system of 
ordinary differential equations (ODEs) -- and allowing discrete mode switching upon transitions. Every transition is equipped with a guard that specifies the switching condition and a reset function that updates the system state after the switch. 

\begin{table}[t]
  \centering
  \caption{Qualitative comparison of different approaches to hybrid system identification. \CmarkB (\XmarkB, resp.) indicates whether the corresponding system feature is supported (or not, resp.); HO is short for high-order.}
  \label{tab:comparison}
  \footnotesize
  %\begin{adjustbox}{max width=.99\linewidth}
  \setlength{\tabcolsep}{1pt}
  \resizebox{1\linewidth}{!}{
  \begin{tabular}{ccccccc}
    \toprule
    \textbf{Method} & \textbf{System Scope} & \textbf{Dynamics Type} & \textbf{HO-ODEs} & \textbf{Resets} & \textbf{Inputs} & \textbf{Noise}\\
    \midrule
    Soto et al.~\cite{garcia2019membership} & Hybrid Automata & Constant ODEs  & \XmarkB & \XmarkB & \XmarkB & \XmarkB\\
    Jin et al.~\cite{DBLP:journals/fac/JinAZZZ21} & Switched Systems & Polynomial ODEs & \XmarkB & \XmarkB & \XmarkB & \XmarkB\\
    Dayekh et al.~\cite{DBLP:conf/cdc/DayekhB024} & Switched Systems & Polynomial ODEs & \XmarkB & \XmarkB & \XmarkB & \XmarkB\\
    POSEHAD \cite{DBLP:journals/tecs/SaberiFB23} & Hybrid Automata & Polynomial ODEs & \XmarkB & \XmarkB & \CmarkB & \XmarkB\\
    LearnHA \cite{DBLP:conf/atva/GurungWS23} & Hybrid Automata & Polynomial ODEs & \XmarkB & \CmarkB & \CmarkB & \XmarkB\\
    HySynth \cite{DBLP:conf/hybrid/SotoH021} & Hybrid Automata & Linear ODEs & \XmarkB & \XmarkB & \XmarkB & \XmarkB\\
    HAutLearn \cite{DBLP:journals/tcps/YangBKJ22} & Hybrid Automata & Linear ODEs & \CmarkB & \XmarkB & \CmarkB & \XmarkB\\
    FaMoS \cite{DBLP:conf/hybrid/PlambeckBHF24} & Hybrid Automata & Linear ARX & \CmarkB & \XmarkB & \CmarkB& \XmarkB \\
    Madary et al.~\cite{DBLP:journals/iandc/MadaryMAL22} & Switched Systems & Nonlinear ARX & \CmarkB & \XmarkB & \CmarkB & \CmarkB\\
    Kochdumper et al.~\cite{kochdumper2025robust} & Hybrid Automata & Polynomial ODEs %Linear ARX/ODEs 
    & \CmarkB & \CmarkB & \CmarkB & \CmarkB\\
    \ourmethod (ours) & Hybrid Automata & Nonlinear ARX & \CmarkB & \CmarkB & \CmarkB & \CmarkB\\
    \bottomrule
  \end{tabular}
  }
  %\end{adjustbox}

\end{table}

This paper addresses the problem of \emph{inferring a hybrid automaton $\haLearned$ from a set of observations $\data$ (i.e., discrete-time traces) of a hybrid system $\ha$}. In particular, $\haLearned$ is expected to closely approximate the behaviors of $\ha$ -- not only on $\data$, but also on unseen traces. To tackle this problem, extensive results have been established in the literature, yet are confined to different subclasses of hybrid systems due to their inherent complexity. For instance, HAutLearn \cite{DBLP:journals/tcps/YangBKJ22} is dedicated to learning hybrid automata with high-order dynamics captured by \emph{linear} ODEs, whilst LearnHA \cite{DBLP:conf/atva/GurungWS23} aims to infer automata with \emph{first-order} polynomial ODEs. See \cref{tab:comparison} for a qualitative comparison of different approaches. A more thorough review of related work is provided in \cref{sec:related-work}.
% \cmscomment{Argue why it's challenging/non-trivial to extend existing approaches, particularly RIHAND~\cite{kochdumper2025robust}, to tackle our setting.}

We present an inference approach called {\ourmethod} to identify nonlinear hybrid automata where the dynamics are captured by \emph{nonlinear autoregressive exogenous (NARX) models} -- a commonly used formalism in hybrid system identification \cite{billings2013nonlinear,DBLP:conf/cdc/LauerBV10}. {\ourmethod} follows the conventional pipeline of trace segmentation -- segment clustering -- mode characterization -- guard and reset learning. However, in contrast to existing approaches that rely heavily on derivative variation and trace similarity for the first two steps, {\ourmethod} works in a \emph{derivative-agnostic} manner by employing NARX models as a unified, threshold-free representation through multiple steps, and thereby unleashes the pipeline for its general applicability to inferring hybrid system models featuring \emph{high-order nonlinear dynamics with exogenous inputs, nonlinear guard conditions, and linear resets} (cf.\ \cref{tab:comparison}). 
%{\color{red} With an extension to \ourmethod, a faithful inference of hybrid systems can be achieved in noisy situations.} 
% Experimental results on a suite of benchmarks demonstrate \ourmethod's distinct superiority in accuracy, applicability, and efficiency compared to state-of-the-art inference tools, as well as its robustness comparable to noise-oriented approaches.
Experimental results on a suite of benchmarks show that \ourmethod significantly outperforms state-of-the-art inference tools in accuracy, applicability, and efficiency, and offers robustness on par with noise-oriented approaches.

\paragraph{\bf Contributions}
Our main contributions are as follows.
\begin{itemize}
    \item We present the {\ourmethod} framework leveraging NARX model fitting for inferring nonlinear hybrid systems 
    %{\color{red} in noiseless/noisy situations}. 
    To the best of our knowledge, {\ourmethod} is the first approach that admits high-order non-polynomial dynamics with non-polynomial inputs/guards, and linear resets.
    \item We show that, unlike threshold-based methods, {\ourmethod} provides formal guarantees on the correctness of both mode-switching detection and trace-segment clustering.
    \item We have implemented {\ourmethod} and conducted extensive experiments to demonstrate its effectiveness, efficiency, and robustness in comparison to state-of-the-art inference tools.
\end{itemize}

% \paragraph{Paper Structure}
% TBA. Additional proofs and experimental details are found in the appendices.
% !TEX root = main.tex

\begin{figure}[t]
    \centering
    \resizebox{1\linewidth}{!}{
    \begin{tikzpicture}
        \draw (0, 0) node[inner sep=0] {\includegraphics[width=1\linewidth]{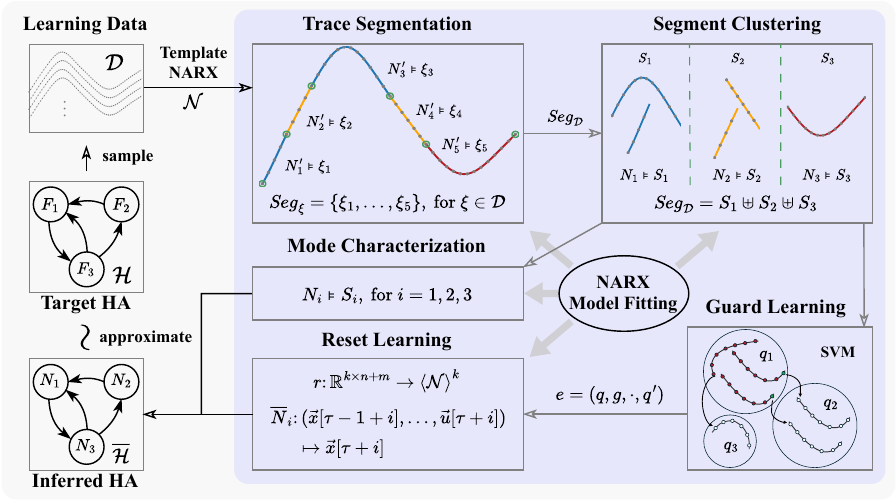}};
        % \draw (1.67, 3.56) node {\small \ourmethod};
        % \draw (2.24, 4.46) node {\large \ourmethod};
        \draw (1.06, 2.13) node {\footnotesize \ourmethod};
    \end{tikzpicture}
    }
    \caption{The general workflow of {\ourmethod}.}
    \label{fig:workflow}
\end{figure}

\section{A Bird's-Eye Perspective}\label{sec:overview}

This section outlines the general workflow of our {\ourmethod} framework as depicted in \cref{fig:workflow}. Given a possibly high-order nonlinear black-box hybrid system $\ha$, {\ourmethod} aims to infer -- from sampled discrete-time input-output traces $\data$ of $\ha$ -- a hybrid automaton $\haLearned$ with NARX-modeled dynamics (subject to a template $\narx$) that closely approximates $\ha$. The inference process in {\ourmethod} adopts a common pipeline of the following steps: 
\begin{enumerate*}[label=(\roman*)]
\item \emph{trace segmentation} for partitioning every trace $\traceDT \in \data$ into a set $\segments_{\traceDT}$ of segments pertaining to different system modes;
\item \emph{segment clustering} for grouping all trace segments that are deemed to follow the same mode dynamics into a cluster $S_i$;
\item \emph{mode characterization} for learning the mode dynamics characterized by a NARX model $\narxIns_i$ under template $\narx$ for each cluster;
\item \emph{guard learning} and \emph{reset learning} for inferring the guard and reset conditions along discrete transitions, respectively.
\end{enumerate*}
The main technical novelty of {\ourmethod} is that it employs \emph{NARX model fitting} as an efficient engine throughout these steps and yields threshold-free inference of a high-order nonlinear hybrid automaton $\haLearned$ that closely approximates the target system $\ha$.

\begin{figure*}[t]
    \centering
    % \begin{subfigure}[b]{0.286\textwidth}
    \begin{subfigure}[b]{0.25\textwidth}
        \centering
        \resizebox{\linewidth}{!}{
        \begin{tikzpicture}[node distance=2cm, auto]
        \node[draw, rounded corners, text width=4cm, align=center, font=\footnotesize] (q1) 
        {$q_1$ (high damping) \\ $x^{(2)} = u - 0.5 x^{(1)} +x - 1.5 x^3$};
        \node[draw, rounded corners, text width=4cm, align=center, font=\footnotesize, below=2cm of q1] (q2) 
        {$q_2$ (low damping) \\ $x^{(2)} = u - 0.2 x^{(1)} +x - 0.5 x^3$};
        \draw[->, bend right=20, -latex', font=\footnotesize] (q1) to node[midway, left, align=right] 
        {$x^2 \leq 0.64$ \\ $x^{(1)}\! := 0.95 x^{(1)}$} (q2);
        \draw[->, bend right=20, -latex', font=\footnotesize] (q2) to node[midway, right, align=left] 
        {$x^2 \geq 1.44$ \\ $x^{(1)}\! := 0.95 x^{(1)}$} (q1);
        \end{tikzpicture}
        }
        \caption{target system $\ha$}
        \label{fig:duffing-system}
    \end{subfigure}%
    \hspace{3mm}
    % \begin{subfigure}[b]{0.38\textwidth}
    \begin{subfigure}[b]{0.331\textwidth}
        \centering
        \includegraphics[width=\textwidth]{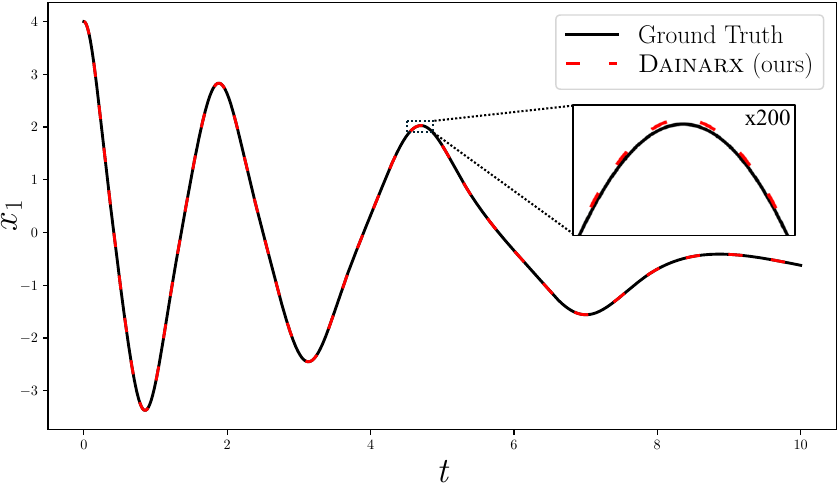}
        \caption{example trace of $\ha$ and $\haLearned$}
        \label{fig:duffing-trace}
    \end{subfigure}
    \hspace{3mm}
    % \begin{subfigure}[b]{0.329\textwidth}
    % \begin{subfigure}[b]{0.288\textwidth}
    \begin{subfigure}[b]{0.25\textwidth}
        \centering
        \resizebox{\linewidth}{!}{
        \begin{tikzpicture}[node distance=5cm, auto]
        \node[draw, rounded corners, text width=4cm, align=center, font=\scriptsize] (q1) 
        {$q_1$ \\ $x[\tau]=2 x[\tau-1]- x[\tau-2]-1.5 \times 10^{-6}x^3[\tau - 1] + 10^{-7}u[\tau]$};
        \node[draw, rounded corners, text width=4cm, align=center, font=\scriptsize, below=1.54cm of q1] (q2) 
        {$q_2$ \\ $x[\tau]=2 x[\tau-1]- x[\tau-2]- 5 \times 10^{-7}x^3[\tau - 1] + 10^{-7}u[\tau]$};
        \draw[->, bend right=20, -latex', font=\scriptsize] (q1) to node[midway, left, align=right] 
        {$\textit{svm}_1$ \\ $r\colon \{\mybar{0.6}{2pt}{N}_1,\mybar{0.6}{2pt}{N}_2$\}} (q2);
        \draw[->, bend right=20, -latex', font=\scriptsize] (q2) to node[midway, right, align=left] 
        {$\textit{svm}_2$ \\ $r\colon \{\mybar{0.6}{2pt}{N}_3,\mybar{0.6}{2pt}{N}_4$\}} (q1);
        \end{tikzpicture}
        }
        \caption{inferred automaton $\haLearned$}
        \label{fig:duffing-result}
    \end{subfigure}
    \caption{Inferring Duffing oscillator via {\ourmethod}. $x^{(i)}$ represents the $i$-th order derivative of $x$; $a := b$ means that $a$ is reset to $b$.}
    \label{fig:oscillator}
\end{figure*}

% {\color{red} The following example demonstrates {\ourmethod}.}

\begin{example}[Duffing Oscillator \textnormal{\cite{KUDRYASHOV2021105526}}]\label{example:duffing}
    The Duffing oscillator is a second-order nonlinear ODE commonly used to describe, e.g., magnetoelastic systems and nonlinear circuits. Consider the hybrid automaton depicted in \cref{fig:duffing-system}, which models a mass undergoing a nonlinear, forced vibration and constantly switches between two controlled modes with different damping coefficients. Each guarded transition between modes is accompanied by a loss of speed (via resets). The input signal $u$ is a cosine function of time $t$ modeling an external driving force. To the best of our knowledge, \emph{none of the existing techniques suffices} to identify a relatively accurate and useful model of this system due to its complex nature of high-order nonlinear dynamics, non-polynomial inputs, and resets.

    Suppose we have collected a dataset $\data$ of 10 discrete-time traces (9 for training, 1 for testing; each with 10,000 data points) via external observations of $\ha$. Then, given a template NARX model with nonlinear terms $x^3[\tau - 1]$ and $x^3[\tau - 2]$ (see details in \cref{subsec:narx-models}), {\ourmethod} first segments the training traces into 155 segments by detecting mode-switching points via NARX model fitting: A model switching is detected if and only if the the current segment under consideration \emph{cannot} be fitted by a single NARX model; Then, all segments that are fittable by the same NARX model are grouped into the same cluster, yielding 2 clusters for this example. The flow dynamics of these two modes are then inferred, again, by NARX model fitting. Finally, we extract the complete hybrid automaton by learning guard conditions using support vector machines (SVMs) with kernel tricks \cite{Boser92} as well as reset functions expressed by NARX models. {\ourmethod} ultimately infers a hybrid automaton $\haLearned$ (\cref{fig:duffing-result}) with NARX-modeled dynamics that closely approximates the behavior of $\ha$: As depicted in \cref{fig:duffing-trace}, the trace generated by $\haLearned$ exhibits a maximum (average, resp.) deviation of 0.0003 ($2.8 \times 10^{-5}$, resp.) from the testing trace of the original system $\ha$. See \cref{sec:experiments} for more details (the \benchmark{duffing} benchmark).
    \qedT
\end{example}
% \znjcomment{The running example is too sketchy, it could be better to provide more explanations, particular, the corresponding five steps should be given.}
% \cmscomment{Revised.}

% !TEX root = main.tex

\section{Problem Formulation}\label{sec:problem-formulation}

% This section first recaps the basics of hybrid systems characterized by hybrid automata and NARX models and then formulates our problem of identifying hybrid systems from time-discrete traces.\cmscomment{May be removed.}

\paragraph*{Notations}
We denote by $\Nats$, $\PosNats$, $\Reals$, and $\PosReals$ the set of natural, positive natural, real, and positive real numbers, respectively; let $\PosRealsInf \defeq \PosReals \cup \{\infty\}$. $X \uplus Y$ denotes the disjoint union of sets $X$ and $Y$. Given matrix $A$, we denote by $A_{ij}$, $A_{i,:}$, and $A_{:,j}$ respectively the $(i,j)$-th element, the $i$-th row vector, and the $j$-th column vector of $A$. For $n \in \PosNats$ and $T \in \PosRealsInf$, an $n$-dimensional (continuous) \emph{signal} $\signal\colon [0,T) \to \RR^n$ is a function that maps each timepoint in the domain $[0,T)$ to an $n$-dimensional real-valued vector in $\RR^n$; a signal is discrete if its time domain consists of discrete timestamps.

\subsection{Hybrid Automata}
Hybrid automata \cite{DBLP:conf/rex/MalerMP91,DBLP:conf/hybrid/AlurCHH92} are a widely adopted formal model for
%formalism to characterize the discrete-continuous behaviors of 
hybrid systems. They extend finite-state machines by associating each discrete location with a vector of continuously evolving dynamics subject to discrete jumps upon transitions.
%\znjcomment{Do you really need to use so general model of HSs? That will complicate your work very much.}
%\cmscomment{I agree. We unfortunately do not have enough time to revise it. Can try in the next version.}
%Such systems are characterized by continuous evolution within discrete modes and abrupt transitions between these modes.
%
%\vspace*{-\baselineskip}
\begin{definition} [Hybrid Automaton \textnormal{\cite{DBLP:conf/rex/MalerMP91,DBLP:conf/hybrid/AlurCHH92}}]
An $n$-dimensional \emph{hybrid automaton} (HA) %(with inputs/outputs) 
is a sextuple 
% $\ha \ \ddefeq\  \!\left(\modes,\, \vars,\, \flows,\, \inv,\, \init,\, \trans\right)$ where 
\begin{equation*}%\label{eq:HA}
    \ha \ \ddefeq\  \!\left(\modes,\, \vars,\, \flows,\, \inv,\, \init,\, \trans\right) \quad \text{where}
\end{equation*}
\begin{itemize}
    \item $\modes = \{q_1, q_2, \ldots, q_j\}$ is a finite set of \emph{discrete modes} representing the discrete states of $\ha$.
    \item $\vars = \varsO \uplus \varsI$ is a finite set of \emph{continuous variables} consisting of (observable) output variables $\varsO = \{x_1, x_2, \ldots, x_n\}$ and (controllable) input variables $\varsI = \{u_1, u_2, \ldots, u_m\}$. A real-valued vector $\vec{v} = (x_1, x_2, \ldots, x_n, u_1, u_2, \ldots, u_m) \in \RR^{n+m}$ denotes a continuous state. The overall \emph{state space} of $\ha$ is $\modes \times \RR^{n+m}$, i.e., the Cartesian product of the discrete and continuous state space. We denote by $\vec{x} = \projO{\vec{v}}$ and $\vec{u} = \projI{\vec{v}}$ the projection of $\vec{v}$ onto the output and input space, respectively.
    % \item $\inputs = \{u_1, u_2, \ldots, u_m\}$ is a finite set of \emph{input variables} with $\vec{u} = (u_1, u_2, \dots, u_m) \in \RR^m$.
    \item $\flows = \{F_q\}_{q\in \modes}$ assigns to each mode $q\in \modes$ a \emph{flow function} $F_q$ characterizing the change of outputs $\varsO$ over inputs $\varsI$; Detailed forms of $F_q$ are specified below.
    \item $\inv \subseteq \modes \times \RR^{n+m}$ specifies the \emph{mode invariants} representing admissible states of $\ha$; We denote by $\inv(q) = \{\vec{v} \mid (q,\vec{v}) \in \inv\}$ the invariant of mode $q \in \modes$.
    \item $\init \subseteq \inv$ is the \emph{initial condition} defining %admissible 
    initial states of $\ha$.
    % \item $\mathcal{G}$ is a set of guard conditions, defining the conditions under which the system switches from one mode to another. Each guard is a set of intervals on the domains of the variables in $\vars$. The guard is considered active if all of the variables are in their given interval.
    \item $\trans \subseteq \modes \times G \times R \times \modes$ is the set of \emph{transition relations} between modes, where $G$ is a set of \emph{guards} $g \subseteq \RR^{n+m}$; $R$ is a set of \emph{resets} (aka, \emph{updates}) $r\colon\RR^{n+m} \to \RR^n$. A transition $e  = (q, g, r, q') \in \trans$ is triggered immediately when $g$ is active\footnote{We assume 
    %all transitions are \emph{deterministic}, i.e., at most one guard (for a source mode) is active at any time.
    \emph{deterministic} transitions where at most one guard is active at any time.}, i.e., $\vec{v} \in g$ (\`{a} la the synchronous semantics in \textnormal{\cite{DBLP:conf/hybrid/NicollinOSY92}}); The output state vector $\vec{x}$ is updated to $r(\vec{v})$ upon taking this transition.
\end{itemize}
\end{definition}%
%\noindent
The flow dynamics of a hybrid automaton $\ha$ within mode $q$ is typically modeled by a system of (potentially high-order) \emph{ordinary differential equations} (ODEs) of the form
\begin{align}\label{eq:odes}
\vec{x}^{(k)} \eeq \,F_q \left(\vec{x}, \vec{x}^{(1)}, \ldots, \vec{x}^{(k-1)},\vec{u}\right)~,
\end{align}%
where $k \in \Nats$ is the \emph{order} of the ODE and $\vec{x}^{(i)} = {\dif^{\,(i)}\!\vec{x}}/{\dif t^i}$ is the $i$-th order derivative of $\vec{x}$ w.r.t.\ time $t$ ($0 \leq i \leq k$). Note that for $k > 1$, the initial condition $\init$ is extended with initial valuations for %the derivatives 
$\vec{x}^{(1)}, \ldots, \vec{x}^{(k-1)}$; these derivatives can also be reset akin to
%in analogous to 
$\vec{x}$.
%We sometimes use $\dot{\vec{x}}$ to denote the first-order time derivative of $\vec{x}$. 

The semantics of a hybrid automaton $\ha$ is usually captured by the set of continuous-time hybrid traces. Such a trace starts from a state
% an initial state 
in $\init$ and evolves according to the flow dynamics $\flows$ (while remaining within invariants $\inv$) and the discrete jumps per $\trans$:
%; see, e.g., \cite{DBLP:journals/siglog/FranzleCK19} for a detailed formulation of such traces.
%
\begin{definition}[Continuous-Time %(Hybrid) 
Trace \textnormal{\cite{DBLP:conf/atva/GurungWS23}}]\label{def:ct-trace}
A \emph{continuous-time hybrid trace} (aka, \emph{run}) of a hybrid automaton $\ha$ with first-order\footnote{We omit the complex yet straightforward extension to 
%the case of 
high-order dynamics ($k >1$).}\! dynamics is a (possibly infinite) sequence 
\begin{equation*}
    \traceCTH \eeq \left\langle \big(q_0, \vec{v}_0\big) \xrightarrow{\tau_0} \big(q_0, \vec{v}'_0\big) \xrightarrow{e_0} \big(q_1, \vec{v}_1\big) \xrightarrow{\tau_1} \big(q_1, \vec{v}'_1\big) \xrightarrow{e_1}  \cdots \right\rangle
\end{equation*}
satisfying %the following conditions:
\begin{enumerate*}[label=(\roman*)]
    \item \emph{initialization:} $(q_0, \vec{v}_0) \in \init$;
    \item \emph{continuous evolution:} for any $i \in \Nats$, there is signal $\signal_i\colon [0,\tau_i] \to \RR^{n+m}$ such that $\signal_i(0) = \vec{v}_i$, $\signal_i(\tau_i) = \vec{v}'_i$. For any $t \in [0,\tau_i]$, $\signal_i(t) \in \inv(q_i)$ and $\signal_i(t)$ satisfies flow dynamics \cref{eq:odes} (with $k=1$)\footnote{This means that ${\dif\vec{x}_i}/{\dif t} = F_{q_i}(\vec{x}_i, \vec{u}_i)$ where $\vec{x}_i$ and $\vec{u}_i$ are the projections of signal $\signal_i$ onto $\varsO$ and $\varsI$, respectively.}; 
    \item \emph{discrete consecution:} for any $e_i = (q_i, g_i,r_i,q_{i+1})$, $\vec{v}'_i \in g_i$ and $\vec{v}_{i+1} = (r_i(\vec{v}'_i), \projI{\vec{v}'_i})$.
\end{enumerate*}
A \emph{continuous-time trace} over run $\traceCTH$ is a signal $\signal\colon [0,T) \to \RR^{n+m}$ such that, $\forall i \in \Nats\colon \signal(t) = \signal_i(t-T_i)$ for $t \in [T_i, T_{i+1})$ with $T_i = \sum^{i-1}_{j=0} \tau_j$ and $T_0 = 0$.
\end{definition}
\noindent

To alleviate \emph{blocking behaviors} of hybrid automata \cite{DBLP:journals/pieee/AsarinBDMP00,DBLP:journals/tac/LygerosJSZS03}, we assume the most permissive form of invariants $\inv = \{(q,\RR^{n+m})\}_{q\in \modes}$ akin to
%along the line of 
\cite{DBLP:conf/atva/GurungWS23,DBLP:conf/hybrid/PlambeckBHF24}.
%\cmscomment{We could also say that our invariants are the complement of guards as in HAutLearn.} 
Moreover, we consider \emph{nonlinear} hybrid automata in the sense that both the \emph{flow dynamics} $\flows$ and the \emph{guard conditions} $G$ can potentially be described by nonlinear functions even beyond polynomials.\footnote{We assume
%allow 
\emph{linear resets} 
%only 
as we use transition-matrix encodings
%representations 
for learning resets.} The duffing oscillation system given in \cref{example:duffing} is a concrete hybrid automaton with second-order ODEs and inputs.

This paper aims to identify hybrid system models in a %commonly adopted paradigm called 
\emph{passive learning} manner %(see, e.g., 
\cite{DBLP:conf/cdc/LauerBV10,DBLP:journals/iandc/MadaryMAL22,DBLP:journals/tcps/YangBKJ22,DBLP:conf/atva/GurungWS23,DBLP:conf/hybrid/PlambeckBHF24},
%), 
i.e., to infer an automaton %$\ha$ 
based on external observations of the system without actively interacting with it during learning. Such observations are typically given by a set of \emph{discrete-time traces}, i.e., sequences of timestamped vectors:
\begin{definition}[Discrete-Time Trace]
A \emph{discrete-time (input-output) trace} of a hybrid automaton $\ha$ with first-order dynamics is a %(possibly infinite)
sequence
\begin{equation*}
    \traceDT \eeq \left\langle \big(t_0,\vec{v}_0\big), \big(t_1,\vec{v}_1\big), \ldots\right\rangle~, \quad \text{with}\ 0 \leq t_0 < t_1 < \cdots
\end{equation*}
such that there exists a continuous-time trace $\signal$ of $\ha$ satisfying $\vec{v}_i = \signal(t_i) \in \RR^{n+m}$ for all $i \in \Nats$. We call such discrete-time trace $\traceDT$ a \emph{discretization} of its continuous-time counterpart $\signal$.
%where $\vec{y}[i] \in \RR^n$ is the observed (under period $\Delta t$) state vector of $\vec{x}(i \cdot \Delta t)$ in $\ha$. A \emph{signal} with respect to $x_i \in \vars$ is the projection of $\traceDT$ onto the corresponding observation $y_i$, denoted by $\signal{y_i}$. 
% \begin{itemize}
%     \item $T$ is the total number of time steps,
%     \item $X_i$ is the values of system variables at time step $i$,
%     \item $t_i$ is the timestamp.
% \end{itemize}
\end{definition}

In practice, given a hybrid system $\ha$, one can obtain a set of \emph{finite} discrete-time traces $\data = \{\traceDT_j\}_{1\leq j \leq M}$ -- called \emph{learning data} -- by \emph{sampling} the observable state $\vec{x}$ under control input $\vec{u}$ of $\ha$. By assuming an identical, fixed sampling period $\Delta t$ for all variables in $V$, we obtain the learning data $\{\traceDT_j\}_{1\leq j \leq M}$ where each finite discrete-time trace $\traceDT_j$ (of length $\traceDTLen+1$) is of the form
\begin{equation}\label{eq:learning-data}
    \traceDT \eeq \left\langle \big(t_0,\vec{v}_0\big), \big(t_0 + \Delta t,\vec{v}_1\big), \ldots, \big(t_0 + \traceDTLen \Delta t,\vec{v}_\traceDTLen\big)\right\rangle~.
\end{equation}
We remark that sampling via external measurements or numerical simulations inevitably introduces \emph{errors}. However, as a convention in system identification, we aim to learn a model that best fits the \emph{sampled data} (while exhibiting predictive power along unseen traces) instead of exact traces of the original system $\ha$ which are often intractable to obtain, especially for nonlinear dynamics \cite{Gan18}.
%Without loss of generality, we assume an identical, fixed sampling period ($\Delta t$) for all variables in $\varsO \cup \varsI$. \cmscomment{Remark on sampling errors?}

To this end, we infer %hybrid system 
models where the flow dynamics $\flows$ is represented using \emph{nonlinear autoregressive exogenous} (NARX) models -- a common formalism used in the identification of (nonlinear) hybrid systems \cite{DBLP:conf/hybrid/LauerB08,DBLP:conf/cdc/LauerBV10,billings2013nonlinear,DBLP:journals/iandc/MadaryMAL22,dayekh2024hybrid,DBLP:conf/hybrid/PlambeckBHF24} -- for fitting discrete-time traces.

% \begin{figure}[t]
% \centering
% \includegraphics[width=2.5in]{f1}
% \caption{Linear hybrid automata for temperature controller.}
% \label{f1}
% \end{figure}

% In Fig. \ref{f1}, we give an example of hybrid automata which models a classic automatic temperature control model \cite{henzinger1996theory}. It has two modes and one variable. The variable $x$ represents the real-time temperature value that changes within the system. When the system stays in the off mode, the heater is turned off and the temperature in the environment decreases according to the flow function $\dot{x} = -0.1x$. When the system is in the on mode, the heater is turned on, and the temperature increases according to the flow function $\dot{x} =5 - 0.1x$. $x>21$ indicate that when the temperature drops below 19°C, the mode can switch from off to on. In contrast, when the temperature increases above 21°C, the mode can be changed back to off. In addition, the model includes two invariants: $x \geq 18$ and $x \leq 22$, which specify the legal range of the variable while in the modes off and on.

\subsection{NARX Models}\label{subsec:narx-models}

We forge the flow dynamics $\flows = \{F_q\}_{q\in \modes}$ using (a system of) NARX models: Each mode dynamics is represented by a \emph{nonlinear} ARX model in the form of a $k$-th order \emph{difference equation}:\footnote{For the sake of simplicity, we consider \emph{noise-free} NARX models. An extended form for fitting \emph{noisy} data is provided in \cref{sec:noisy}. %see \cref{sec:limitations} for a detailed discussion.
}
\begin{equation}\label{eq:narx-standard}
    \vec{x}[\tau] = F_q\left(\vec{x}[\tau - 1], \vec{x}[\tau - 2],\ldots, \vec{x}[\tau - k], \vec{u}[\tau]\right)~, \quad \text{for}\ \tau \geq k
\end{equation}%
where the current output vector $\vec{x}[\tau] \in \RR^n$ is determined by its $k$ recent history states and the current input vector $\vec{u}[\tau] \in \RR^m$ through a nonlinear function $F_q\colon \RR^{k \times n} \times \RR^m \to \RR^n$.

We observe that, in many practical applications, the form of differential dynamics as well as their difference counterparts can be expressed as a linear combination of simple nonlinear terms and linear terms. This gives rise to an equivalent form of %the NARX model 
\cref{eq:narx-standard} as
%\begin{equation}
\begin{multline}\label{eq:narx-templated}
    \vec{x}[\tau] \eeq \sum\nolimits_{i=1}^{\alpha} \vec{a}_i \hadamard %\underbrace{
    f_i\left(\vec{x}[\tau - 1], \ldots, \vec{x}[\tau - k], \vec{u}[\tau]\right)
    %}_{\mathclap{\textnormal{nonlinear terms}}}
    \,+\, \\
    % \underbrace{
    \sum\nolimits_{i=1}^{k} B_i \cdot \vec{x}[\tau - i] \,+\, B_{k+1} \cdot \vec{u}[\tau]%}_{\mathclap{\textnormal{linear terms}}}
    \,+\,
    \vec{c}~,
\end{multline}%
%\end{equation}%
where $f_1, f_2, \ldots, f_\alpha \in \RR^{n}$ are arbitrary vector-valued nonlinear components in $\RR^{n}$
%\emph{kernel functions} \cite{DBLP:conf/cdc/LauerBV10,DBLP:journals/iandc/MadaryMAL22} 
representing the model structure of $F_q$; $\vec{a}_i, \vec{c} \in \RR^{n}$ are vector coefficients and $B_i \in \RR^{n\times n}, B_{k+1} \in \RR^{n\times m}$ are matrix coefficients; $\cdot$ denotes matrix multiplication and $\hadamard$ denotes the Hadamard product, i.e., $(A_1 \hadamard A_2)_{ij} = (A_1)_{ij} (A_2)_{ij}$. For a compact representation, we maintain a \emph{coefficient matrix} $\cmat \in \RR^{n \times (\alpha + k n + m + 1)}$ of \cref{eq:narx-templated}:
\begin{equation*}
    \cmat \eeq \begin{bmatrix}\vec{a}_1 & \vec{a}_2& \ldots& \vec{a}_\alpha& B_1& B_2& \ldots& B_k& B_{k+1}& \vec{c}\end{bmatrix}~,
\end{equation*}%
and thereby denote by $\narx_q = (\cmat, \{f_i\}_{1 \leq i \leq \alpha})$ the flow dynamics of a hybrid system within mode $q$.
% and thereby denote by $\narx_q = \left\langle k, \vars, \inputs, \cmat, \{f_i\}_{1 \leq i \leq \alpha} \right\rangle$ the flow dynamics of a hybrid system within mode $q$.

A \emph{template NARX model} refers to flow dynamics $\narx$ whose nonlinear components $\{f_i\}_{1 \leq i \leq \alpha}$ are given yet the coefficient vector $\cmat$ consists of \emph{unknown parameters}. Let $\cmatIns$ be an instantiation of $\cmat$. We denote by $\narxIns = \narx[\cmatIns]$ the \emph{instance} of $\narx$ under $\cmatIns$ and by $\narxInsSet$ the set of all instances of $\narx$.

A \emph{discrete-time trace} of a hybrid automaton $\ha$ with NARX-modeled dynamics can be defined as a discrete signal $\traceNARX\colon \Nats \to \RR^{n+m}$ in analogy to %(the definition of continuous-time trace in) 
\cref{def:ct-trace}, except that we replace the ODE-modeled dynamics \cref{eq:odes} with the NARX dynamics \cref{eq:narx-templated}.

The identification problem %concerned 
in this paper reads as follows:
\begin{tcolorbox}[boxrule=1pt,colback=white,colframe=black!75,boxsep=0mm]%[boxrule=0pt,colframe=black!75]
\textbf{Problem Statement.}\ \ Given a set of %finite 
discrete-time traces $\{\traceDT_j\}_{1\leq j \leq M}$ %as 
per \cref{eq:learning-data} of
%sampled from 
a hybrid automaton $\ha$ and a template NARX model $\narx = ( \cmat, \{f_i\}_{1 \leq i \leq \alpha} )$, identify a hybrid automaton $\haLearned$ where all mode dynamics are instances of $\narx$ such that, for any trace $\traceDT_j$, there exists a finite discrete-time trace $\traceNARX$ of $\haLearned$ satisfying $\traceNARX(\tau) = \vec{v}_\tau$ for all $0 \leq \tau \leq \traceDTLen$.
\end{tcolorbox}
%
%\noindent
The above problem statement specifies the ideal setting where the inferred model $\haLearned$ \emph{coincides} with the target model $\ha$ over the given learning data. In practice, however, one often aims to learn a model $\haLearned$ that \emph{approximates} $\ha$ in the sense that $\haLearned$ is capable of generating good predictions along both given traces (for training) and unseen traces (for testing) of $\ha$.
% !TEX root = main.tex

\section{Derivative-Agnostic Inference}\label{sec:inference}

This section presents our derivative-agnostic approach {\ourmethod} for inferring hybrid systems with NARX-modeled dynamics from discrete-time input-output traces. As has been exemplified in \cref{sec:overview}, {\ourmethod} follows the common pipeline of hybrid system identification including
\begin{enumerate*}[label=(\Roman*)]
\item \emph{Trace segmentation}: It partitions every discrete-time trace in the learning data into trace segments by the detection of changepoints, namely, data points on the trace that signify a change of dynamics (i.e., mode switching) between neighboring segments; \label{step:segmentation}
\item \emph{Segment clustering}: It groups the trace segments into several clusters such that each cluster includes all segments that are deemed to be generated by the same mode dynamics; \label{step:clustering}
\item \emph{Mode characterization}: It learns a NARX model adhering to the given template to characterize the mode dynamics for each cluster; \label{step:dynamics}
\item \emph{Guard learning}: It learns a guard condition for each mode switching using SVMs; \label{step:guard-learning}
\item \emph{Reset learning}: It learns a reset function for each mode switching again using NARX models. \label{step:reset-learning}
\end{enumerate*}

The main technical novelty of {\ourmethod} lies in \cref{step:segmentation,step:clustering,step:reset-learning}; In particular, we employ \emph{NARX model fitting} as the basis throughout Steps 
\ref{step:segmentation}, \ref{step:clustering}, \ref{step:dynamics}, and \ref{step:reset-learning}, which aims to fit one or more trace segments using a NARX model. In the rest of this section, we first present NARX model fitting, then elaborate each step of {\ourmethod}, and finally provide an algorithmic complexity analysis. 

\subsection{NARX Model Fitting}\label{subsec:narx-fitting}

Given a template NARX model $\narx = ( \cmat, \{f_i\}_{1 \leq i \leq \alpha} )$ and a set $S = \{\traceDT_j\}_{1\leq j \leq M}$ of %finite 
discrete-time traces or trace segments (cf.\ \cref{subsec:segmentation}) where each trace (segment) takes the form $\traceDT = \langle (t_0,\vec{v}_0), (t_0 + \Delta t,\vec{v}_1), \ldots, (t_0 + \traceDTLen \Delta t,\vec{v}_\traceDTLen)\rangle$, the task of \emph{NARX model fitting} asks to find an instance $\narxIns = \narx[\cmatIns] \in \narxInsSet$, if it exists, such that $\narxIns$ \emph{fits} $S$:
\begin{definition}[Trace Fitting]
    \label{def:trace-fitting}
    Let $\traceDT$ be a finite discrete-time trace of hybrid automaton $\ha$. We say that a $k$-th order NARX model $\narxIns$ \emph{fits} $\traceDT$, denoted by $\narxIns \models \traceDT$, if \cref{eq:narx-templated} holds by substituting $\projO{\vec{v}_{\tau}}$ and $\projI{\vec{v}_{\tau}}$ in $\traceDT$ respectively for $\vec{x}[\tau]$ and $\vec{u}[\tau]$. $\narxIns$ fits a set $S = \{\traceDT_j\}$ of such trace (segments) if $\narxIns \models \traceDT_j$ for all $\traceDT_j$ in $S$. We say $S$ is \emph{fittable} by a template NARX model $\narx$ if there exists $\narxIns \in \narxInsSet$ such that $\narxIns \models S$.
\end{definition}

We start with the simple case of fitting a \emph{single} trace %(segment) 
$\traceDT$ by finding $\cmatIns \in \cmat$ such that $\narxIns = \narx[\cmatIns] \models \traceDT$. We address this task by the \emph{linear least squares method} (LLSQ) as standard in regression analysis \cite{DBLP:books/ox/07/GolubR07}: We find the best-fitting NARX coefficients $\cmatIns$ to $\traceDT$ by minimizing the sum-of-squares of the \emph{offsets of data points} on $\traceDT$ from $\narx[\cmatIns]$: %i.e.,
\begin{mini}|l|
    {\cmatIns_{i,:}}
    {\normDisplay{\Omatrix_{i,:} - \cmatIns_{i,:} \cdot \Dmatrix_i} \quad \text{for}\ i = 1, 2, \ldots, n}
    {\label{eqn:optimization}}
    {}
    % \addConstraint{M + M_2}{\ssucceq 0~}{}
    % \addConstraint{M_2}{\ssucceq 0~.}{}
\end{mini}%
where $\norm{\cdot}$ is the $L^2$-norm; $\Dmatrix_i \in \RR^{(\alpha + k n + m + 1) \times (\traceDTLen - k + 1)}$ and $\Omatrix \in \RR^{n \times (\traceDTLen - k + 1)}$ are respectively the \emph{data matrix} (w.r.t. $i$-th row of $\cmatIns$) and the \emph{observed value matrix}, which are constructed as follows. Let $\vec{x}_{\tau} = \projO{\vec{v}_\tau}$ and $\vec{u}_{\tau} = \projI{\vec{v}_\tau}$ for $0 \leq \tau \leq \traceDTLen$, then the observed value matrix $\Omatrix$ is
% Assuming there exists an instance of $\narx$ called $\narx' = \langle \mathcal{A}, \{f_i\}_{1 \leq i \leq \alpha} \rangle$, our task is equivalent to obtaining instance $\mathcal{A}$ of $\cmat$. We need to talk about each line of $\mathcal{A}$ separately. For the i-th row $\mathcal{A}_{i,:}$ of $\mathcal{A}$, a data matrix $Q_i \in \RR^{(\alpha + kn + m + 1) \times (\traceDTLen - k + 1)}$ is constructed:
$\Omatrix =
\begin{bmatrix}
  {\vec{x}_{\traceDTLen}} \!&\! {\vec{x}_{\traceDTLen - 1}} \!&\! \cdots \!&\! {\vec{x}_{k}}
\end{bmatrix} %\quad \text{with}\ \vec{x}_{j} = \projO{\vec{v}_j} \text{for}\ 0 \leq j \leq \traceDTLen;
$;
% \begin{equation*}
% \Omatrix \eeq
% \begin{bmatrix}
%   {\vec{x}_{\traceDTLen}} & {\vec{x}_{\traceDTLen - 1}} & \cdots & {\vec{x}_{k}}
% \end{bmatrix}~; %\quad \text{with}\ \vec{x}_{j} = \projO{\vec{v}_j} \text{for}\ 0 \leq j \leq \traceDTLen;
% \end{equation*}%
and the data matrix $\Dmatrix_i$ is %constructed as
\begin{gather*}
    \Dmatrix_i \eeq
    \begin{bmatrix}
    \chi_{i,\traceDTLen} & \chi_{i,\traceDTLen - 1} & \dots & \chi_{i,k} 
    \end{bmatrix} \quad \text{where}\\
% \end{equation*}%
% \begin{equation*}
% \begin{aligned}
    \chi_{i,\tau} =
    \begin{bmatrix}
    f_{1}(\vec{y}_\tau)_{i,:} \!\!&\!\! \cdots \!\!&\!\! f_{\alpha}(\vec{y}_\tau)_{i,:} \!\!&\!\! \vec{y}_\tau^\transpose \!\!&\!\! 1
    \end{bmatrix}^\transpose \text{with}\,\, \vec{y}_\tau = \left({\vec{x}_{\tau - 1}}, \cdots, {\vec{x}_{\tau - k}}, {\vec{u}_{\tau}}\right)\,.
% \end{aligned}
\end{gather*}%

The linear optimization problem \cref{eqn:optimization} can be discharged using any off-the-shelf solvers by finding its least-squares solution \cite{DBLP:books/ox/07/GolubR07}. The following example illustrates the optimization problem:
\begin{example}[Trace Fitting via LLSQ]
    To fit the trace $\traceDT = \left\langle \big(t_0,\vec{v}_0\big)\right.\allowbreak \left.\big(t_0 + \Delta t,\vec{v}_1\big), \ldots, \big(t_0 + 4 \Delta t,\vec{v}_4\big)\right\rangle$ under the second-order \emph{univariate} template NARX model $\narx = (\cmat, \{x^2[\tau - 1], x^3[\tau - 2]\})$ (where all coefficients in $\cmat$ are of dimension $1 \times 1$), the matrices are constructed as $\Dmatrix_1 = \left[\, x_3^2 ,\, x_2^3 ,\, x_3 ,\, x_2 ,\, 1 ;\,  x_2^2 ,\, x_1^3 ,\, x_2 ,\, x_1 ,\, 1 ;\, x_1^2 ,\, x_0^3 ,\, x_1 ,\, x_0 ,\, 1\,\right]^\transpose$ and $\Omatrix = \left[ \, x_4 ,\, x_3 ,\, x_2 \, \right]$.
    % $\Dmatrix_1 = \left( \begin{smallmatrix} x_3^2 & x_2^2 & x_1^2 \\
    %         x_2^3 & x_1^3 & x_0^3 \\
    %         x_3 & x_2 & x_1 \\
    %         x_2 & x_1 & x_0 \\
    %         1 & 1 & 1\end{smallmatrix} \right)$
    % \begin{equation}
    % \nonumber
    %     \Omatrix = \begin{bmatrix}
    %         x_4 & x_3 & x_2
    %     \end{bmatrix}
    % \end{equation}
    % \begin{equation}
    % \nonumber
    %     \Dmatrix_1 = \begin{bmatrix}
    %         x_3^2 & x_2^2 & x_1^2 \\
    %         x_2^3 & x_1^3 & x_0^3 \\
    %         x_3 & x_2 & x_1 \\
    %         x_2 & x_1 & x_0 \\
    %         1 & 1 & 1
    %     \end{bmatrix}
    % \end{equation}
    %
    Then, the %optimization 
    objective in \cref{eqn:optimization}, for the first column, is $x_4 - (a_1 \cdot x_3^2 + a_2 \cdot x_2^3 + B_1 \cdot x_3 + B_2 \cdot x_2 + c \cdot 1)$ (cf.\ \cref{eq:narx-templated}).
    \qedT
\end{example}

% From the matrix $Q_i$ and the row vector $P_i$, we can write the corresponding form of \cref{eq:narx-templated} on the i-th variable: $P_{i,:} = \mathcal{A}_{i.:} \times Q_i$. $\mathcal{A}_{i.:}$ can therefore be obtained by the following optimization objectives
% \begin{equation}
%     \label{Coefficient optimization}
%     \min \left \| P_{i,:} - \mathcal{A}_{i.:} \times Q_i\right \| 
% \end{equation}
% Using least square method, we can obtain the row vector $\mathcal{A}_{i,:}$, and the coefficient matrix $\mathcal{A}$.

The above LLSQ-based fitting can be generalized to the case of multiple trace (segments) with $S = \{\traceDT_j\}_{1\leq j \leq M}$ by extending the matrices $\Omatrix$ and $\Dmatrix_i$ as
\begin{equation}
% \begin{aligned}
    \label{eq:extended-matrices}
        \Omatrix \eeq \begin{bmatrix}
        \Omatrix^{\traceDT_1} \; 
        \Omatrix^{\traceDT_2} \; 
        \cdots \;
        \Omatrix^{\traceDT_M}
        \end{bmatrix} \ \ \text{and}\ \ 
        \Dmatrix_i \eeq \begin{bmatrix}
        \Dmatrix^{\traceDT_1}_i \;
        \Dmatrix^{\traceDT_2}_i \;
        \cdots \;
        \Dmatrix^{\traceDT_M}_i \;
        \end{bmatrix}~,
% \end{aligned}
\end{equation}%
where $\Omatrix^{\traceDT_j}$ and $\Dmatrix_i^{\traceDT_j}$ are %the 
corresponding matrices constructed for %trace 
$\traceDT_j$.

% The same method can also be extended to simultaneously fit multiple traces. Assuming we have a set of discrete-time traces $S = \{\traceDT_j\}_{1\leq j \leq M}$, and a template NARX model $\tilde{\narx}$, our goal is to find an instance $\narx$ that satisfies $\narx \models S$

% We only need to concatenate the $Q_i$ matrix and the $P$ matrix from the previous method into the following form, respectively.

% % Where $..$ and P are respectively D_i and P matrices obtained by i through the above method
% % Recompute the \(\mathcal{A}\) corresponding to Equation \ref{Coefficient optimization} to obtain \(\mathcal{A}_S\) and \(E_S\).

It follows that the NARX model $\narx[\cmatIns]$ obtained by discharging the optimization problem \cref{eqn:optimization} with \cref{eq:extended-matrices} indeed fits the set of trace (segments) $S$ if the optimal value is zero for all $1 \leq i \leq n$: 
%We say $\{\traceDT_j\}$ is \emph{fittable} by a template NARX model $\narx$ if $E_{\{\traceDT_j\}} = 0$.
%
\begin{restatable}[Correctness of Trace Fitting]{theorem}{restateExactFitting}
\label{thm:exact-fitting}
% \begin{theorem}[Correctness of Trace Fitting]\label{thm:exact-fitting}
Given a template NARX model $\narx$ and a set of finite discrete-time trace (segments) $S = \{\traceDT_j\}_{1\leq j \leq M}$. Let $E_S = \max \{E_1, E_2, \ldots, E_n\}$ where each $E_i$ is the optimal value of \cref{eqn:optimization} with \cref{eq:extended-matrices} w.r.t.\ $i$; let $\narxIns = \narx[\cmatIns] \in \narxInsSet$ be the NARX model corresponding to the optimum $\cmatIns$. Then, $S$ is fittable by $\narx$ if and only if $E_S = 0$. In this case, we have $\narxIns \models S$.
% \end{theorem}
\end{restatable}

% In order to verify whether the NARX model $\narx$ formed by the coefficient matrix obtained by this method can fit the discrete-time trace $\traceDT$, we define the deviation value:
% \begin{equation}
%     \begin{split}
%         E_i &= \max \left | P_{i,:} - \mathcal{A}_{i.:} \times Q_i \right | \\
%         E_{\traceDT} &= \max \{ E_1, E_2, \dots, E_n\}
%     \end{split}
% \end{equation}

\subsection{Trace Segmentation}\label{subsec:segmentation}

Given learning data $\data = \{\traceDT_j\}_{1\leq j \leq M}$, {\ourmethod} first partitions
%the initial step of {\ourmethod} is to partition 
every discrete-time trace in $\data$ into trace segments such that the neighboring segments are generated by different mode dynamics. Intuitively, a \emph{trace segment} is a finite subsequence of a discrete-time trace:
%by the detection of changepoints, namely, data points on the trace that signify a change of dynamics (i.e., mode switching) between neighboring segments;
% The goal of trace segmentation is to separate the trace segments of different modes in the entire trace. We first define the trace segment:
%
\begin{definition}[Trace Segment]\label{def:trace-segment}
    Given a finite discrete-time trace $\traceDT = \langle (t_0,\vec{v}_0), (t_0 + \Delta t,\vec{v}_1), \ldots, (t_0 + \traceDTLen \Delta t,\vec{v}_\traceDTLen)\rangle$ and $l,h \in \Nats$ satisfying $l < h \leq \traceDTLen+1$. The segment of $\traceDT$ over $[l,h)$ is $\traceDT_{l,h} \defeq \langle (t'_0,\vec{v}'_0), (t'_0 + \Delta t,\vec{v}'_1), \ldots, (t'_0 + (h - l -1) \Delta t,\vec{v}'_{h - l - 1})\rangle$, where $t'_0 = t_0 + l \Delta t$ and for any $\tau < h - l$, $\vec{v}'_\tau=\vec{v}_{\tau + l}$.
\end{definition}%

By definition, a trace segment is also a (finite) discrete-time trace. Since the segmentation procedure is identical for every trace in the learning data $\data$, 
we consider an individual trace $\traceDT \in \data$. Recall that the semantics of hybrid systems considered in this paper is deterministic and \emph{synchronous} \cite{DBLP:conf/hybrid/NicollinOSY92}: A transition of a hybrid automaton is triggered \emph{immediately} upon the activation of its guard and thus will be reflected by the corresponding mode switching in the observed discrete-time traces. Existing derivative-based segmentation approaches -- e.g., FaMoS \cite{DBLP:conf/hybrid/PlambeckBHF24} which uses linear ARX models for trace fitting -- determine mode switching (and thus a segmentation point) based on the presence of large deviations between immediate future and immediate past measured by sliding windows \cite{DBLP:journals/sigpro/TruongOV20} around the point. In {\ourmethod}, we also employ a sliding window along the trace, however, the mechanism for detecting mode switching is essentially different:
% employes an essentially different
% %, we also employ a sliding window along the trace, however, 
% mechanism for detecting mode switching: 
We identify a set of \emph{timestamps} featuring the following \emph{changepoint property}:
\begin{property}[Changepoints]\label{property:changepoints}
Given a discrete-time trace $\traceDT = \langle (t_0,\vec{v}_0)$, $ (t_0 + \Delta t,\vec{v}_1), \ldots, (t_0 + \traceDTLen \Delta t,\vec{v}_\traceDTLen)\rangle$ and a template NARX model $\narx$. Let $\cpoints = \{p_0, p_1,\ldots,p_s\} \in \Nats^s$ be a set of timestamps with $0 = p_0 < p_1 < p_2 < \cdots < p_s = \traceDTLen + 1$. We say $\cpoints$ is the (unique) set of \emph{changepoints} w.r.t.\ $\traceDT$ and $\narx$ if it satisfies
\begin{enumerate}[label=(\roman*)]
    \item $\traceDT_{p_i,p_{i+1}}$ is \emph{fittable} by $\narx$ for any $0 \leq i < s$; and
    \item $\traceDT_{p_i,(p_{i + 1})+1}$ is \emph{not fittable} by $\narx$ for any $0 \leq i < s - 1$.
\end{enumerate}
\end{property}

{\makeatletter
\let\par\@@par
\par\parshape0
\everypar{}
\begin{wrapfigure}{r}{0.5\linewidth}
\vspace*{-5mm}
    \centering
    \begin{tikzpicture}
        \draw (0, 0) node[inner sep=0] {\includegraphics[width=.98\linewidth]{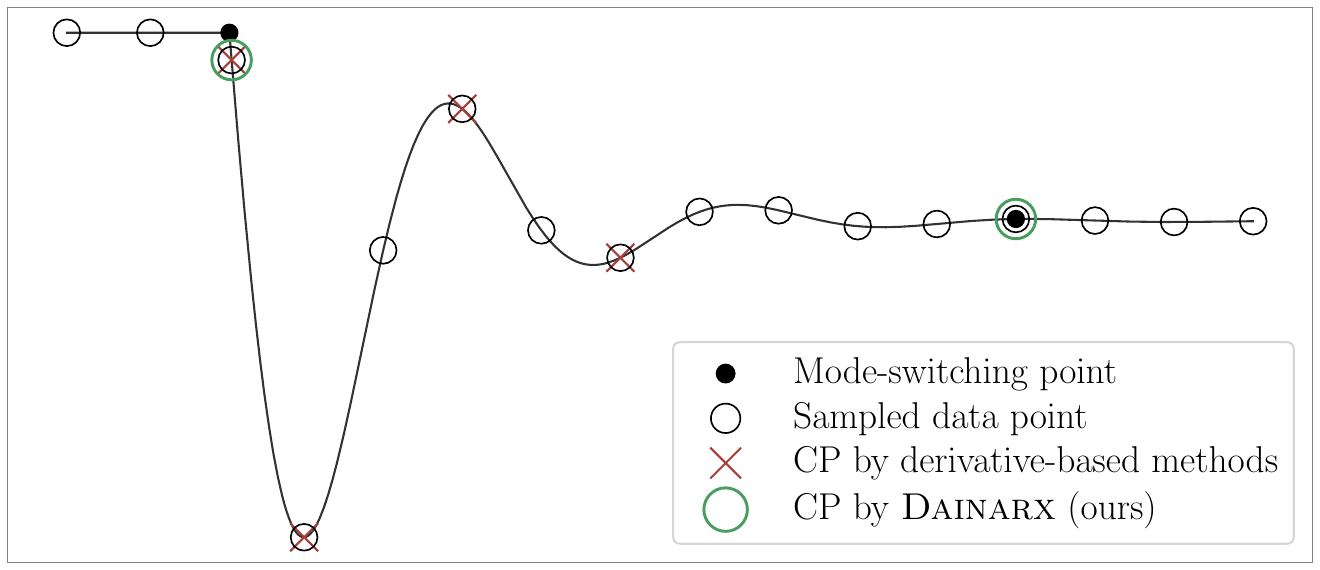}};
        \draw (-1.38, -.74) node {\tiny $\traceDT(3)$};
        \draw (-.64, .73) node {\tiny $\traceDT(5)$};
        \draw (-0.1, -.12) node {\tiny $\traceDT(7)$};
        \draw (1.16, .4) node {\tiny $\traceDT(12)$};
    \end{tikzpicture}
    \vspace*{-3.2mm}
    \caption{Intuition of segmentation.}
    \label{fig:segmentation}
\vspace*{-3.5mm}
\end{wrapfigure}
\noindent
Based on the identified set $\cpoints$ of changepoints, we can obtain the set of segments of trace $\traceDT$ as 
        $\segments_{\traceDT} = \{\traceDT_{p_0, p_1}, \traceDT_{p_1, p_2}, \ldots, \traceDT_{p_{s-1}, p_s}\}$.
\\For the entire learning data 
\par}
\noindent
$\data = \{\traceDT_j\}_{1\leq j \leq M}$, the set of all trace segments is $\segments_{\data} = \bigcup_{j=1}^{M} \segments_{\traceDT_j}$. The intuition behind segmentation in {\ourmethod} against derivative-based segmentation is depicted in \cref{fig:segmentation}. We note that, in FaMoS, the presence of \enquote{large deviations} depends on a \emph{user-supplied threshold} $\gamma$, which is difficult to provide in practice since genuine mode-switching points may be missed (e.g., $\traceDT(12)$ in \cref{fig:segmentation}) if $\gamma$ is too large while spurious mode-switching points may be identified (e.g., $\traceDT(3), \traceDT(5), \traceDT(7)$ in \cref{fig:segmentation}) if $\gamma$ is too small. See example in \cref{subsec:experiments-seg-clustering}.

% \begin{figure}[!t]
%     \centering
%     \resizebox{.6\linewidth}{!}{
%     \begin{tikzpicture}
%         \draw (0, 0) node[inner sep=0] {\includegraphics[width=.98\linewidth]{paper/fig/seg1.pdf}};
%         \draw (-2.64, -1.56) node {\footnotesize $\traceDT(3)$};
%         \draw (-1.27, 1.4) node {\footnotesize $\traceDT(5)$};
%         \draw (-0.25, -.13) node {\footnotesize $\traceDT(7)$};
%         \draw (2.22, .75) node {\footnotesize $\traceDT(12)$};
%     \end{tikzpicture}
%     }
%     %\vspace*{-1.8mm}
%     \caption{Segmentation: {\ourmethod} vs.\ derivative-based strategies.}
%     \label{fig:segmentation}
% \end{figure}

% {\makeatletter
% \let\par\@@par
% \par\parshape0
% \everypar{}
% \begin{wrapfigure}{r}{0.48\linewidth}
% \vspace*{-1mm}
%     \centering
%     \begin{tikzpicture}
%         \draw (0, 0) node[inner sep=0] {\includegraphics[width=.98\linewidth]{paper/fig/seg.pdf}};
%         \draw (-2.2, -1.86) node {\tiny $\traceDT(3)$};
%         \draw (-.98, 1.54) node {\tiny $\traceDT(5)$};
%         \draw (0, -.06) node {\tiny $\traceDT(7)$};
%         \draw (1.84, .74) node {\tiny $\traceDT(12)$};
%     \end{tikzpicture}
%     \vspace*{-1.8mm}
%     \caption{Segmentation: {\ourmethod} vs.\ derivative-based strategies.}
%     \label{fig:segmentation}
% %\end{center}
% \vspace*{-4.5mm}
% \end{wrapfigure}
% \noindent  points may be identified (e.g., $\traceDT(3), \traceDT(5), \traceDT(7)$ in \cref{fig:segmentation}) if $\gamma$ is too small. See example in \cref{subsec:experiments-seg-clustering}.
% \par}

In order to identify the set of changepoints as per \cref{property:changepoints}, we adopt a \emph{sliding window} (of size $\windowSize > k+1$) along every trace $\traceDT \in \data$; the details are given in \cref{alg:segmentation-sliding}. We remark that, in Lines \ref{alg:segmentation-sliding-check} to \ref{alg:segmentation-sliding-drop}
%\cmscomment{The references should be \enquote{Lines 9 to 11}. But I have no clue why it produces \enquote{Algorithm 1}.} 
of \cref{alg:segmentation-sliding}, we perform a \emph{posterior screening} to ensure that every trace segment in the eventually obtained set $\segments_{\traceDT}$ is fittable by the template NARX model $\narx$. Such a screening process is necessary due to the following three 
\emph{corner cases}: (i) the given template NARX model 
$\narx$ does not suffice to fit some trace segment(s) (within a mode);
(ii) the distance between some neighboring mode-switching points falls below the size $\windowSize$ of the sliding window;
(iii) two segments $\traceDT$ and $\traceDT'$ with $\windowSize-1$ overlapped data points are both fittable by $\narx$, yet $\{\traceDT, \traceDT'\}$ is \emph{not} fittable by $\narx$ -- in this case, we may miss the changepoint in $\traceDT'$.
In the absence of these corner cases (see details in the proof of \cref{thm:segmentation-sliding}, \ifarxiv\cref{sec:appendix-proofs}\else\cite[Appx.~A]{DBLP:journals/corr/abs-2507-16426}\fi), we can establish the correctness of \cref{alg:segmentation-sliding} as follows:
%
% \vspace*{-1mm}
\begin{restatable}[Correctness of Segmentation]{theorem}{restateSegmentation}
%in \cref{alg:segmentation-sliding}
\label{thm:segmentation-sliding}
% \begin{theorem}[Correctness of Segmentation]\label{thm:segmentation-sliding}
Given a discrete-time trace $\traceDT$ and a template NARX model $\narx$, the set $\segments_{\traceDT}$ of trace segments and the corresponding set of changepoints $CP$ returned by \cref{alg:segmentation-sliding} -- in the absence of the above three corner cases -- satisfies the changepoint property in \cref{property:changepoints}.
% \end{theorem}
\end{restatable}

% % Since this task is independent for different traces, we discuss algorithms for each individual trace $\traceDT$. We refer to the previous work and use the sliding-window method. 
% Assuming that the window size is $\windowSize$, move the left end of the window from $0$ to $l - \windowSize + 1$ in order, and according to the method mentioned in \cref{subsec:narx-fitting}, we can find out whether the trace segment of the window is fittable by $\narx$. Assuming that the left end of the current window is located at $l > 0$, if there is $\traceDT_{l - 1,l + \windowSize - 1}$ that is fittable by $\narx$, but $\traceDT_{l,l + \windowSize}$ is not fittable by $\narx$, $l + \windowSize - 1$ is a mode change point, so add $l + \windowSize - 1$ to the change points sequence $\cpoints$ and jump $\windowSize$ to the right of the window. 

% Finally, in order to ensure the correctness of trace segmentation, we will check whether each segment can is fittable by $\narx$. If the verification fails, the segment is marked as invalid to prevent subsequent processes from being affected. Therefore, this method can guarantee that the obtained mode change points are in line with the goal under the condition that the interval of the real mode change points is greater than $\windowSize$ and no invalid segments are found.

% \vspace*{-2mm}
% \begin{comment}
\begin{remark}
    We can replace the sliding window-based search of changepoints in \cref{alg:segmentation-sliding} with a \emph{binary search}. The so-obtained algorithm (\ifarxiv see \cref{sec:appendix-algorithms}\else\cite[Appx.~B]{DBLP:journals/corr/abs-2507-16426}\fi) exhibits a (logarithmic) increase in time complexity, yet yields stronger theoretical guarantees (cf.\ \cref{thm:segmentation-binary}). 
    % Nonetheless, we observe that, in practice, the binary search-based segmentation relies on high-quality template NARX models. Therefore, we opt for the sliding window-based implementation, which yields considerably accurate models even in the absence of high-quality templates, as demonstrated in \cref{subsec:efficiency}.
    % %Since efficiency is not the primary factor considered in offline learning, we opt for the binary search-based implementation which already exhibits competitive efficiency in practice as demonstrated in \cref{subsec:efficiency}.
\end{remark}
% \end{comment}

% In addition to the sliding-window method, we also design a binary-based trace segmentation method, which sacrifices a small amount of efficiency and can guarantee the correctness of the algorithm in more cases. The pseudocode for this method is given in the Appendix \cite{?}. \mbhcommentinline{Add binary based algorithm}

\subsection{Segment Clustering} \label{subsec:clustering}

% \begin{definition}[Changepoint]
% Changepoints are a sequence of timestamp defined as:
% \begin{equation}
%     [\tau_i]_{i=0,1,\dots,p} \subseteq T
% \end{equation}
% \end{definition}
% A changepoint $\tau_i$ must satisfy the following conditions:(1) $\tau_0 =0,\tau_i<\tau_{i+1}$ and $\tau_p =T$, (2) $\forall{i}\forall{t}\in[\tau_i,\tau_{i+1}),q_t=q_i$, (3) $\forall{i}\forall{t}\in[\tau_i,\tau_{i+1})$, the flow function $\flows_t$ at each $t$ is the same as at $\tau_i$.
Segment clustering aims to group trace segments into several clusters such that each cluster contains all segments deemed to be generated by the same mode dynamics. The key to this procedure is the criterion for merging a set of trace segments into the same cluster. Existing approaches, such as POSEHAD \cite{DBLP:journals/tecs/SaberiFB23}, FaMoS \cite{DBLP:conf/hybrid/PlambeckBHF24}, and LearnHA \cite{DBLP:conf/atva/GurungWS23}, mainly adopt the criterion based on \emph{trace similarity} computed by, e.g., the dynamic time warping (DTW) algorithm \cite{berndt1994using}, which again depend on \emph{user-supplied thresholds} to determine the clustering,
%of segments, 
and may induce wrong results because \enquote{similar traces} do not imply \enquote{same mode dynamics}, neither vice versa (for different initial values); see examples in \cref{subsec:experiments-seg-clustering}. In contrast, {\ourmethod} employs a \emph{threshold-free} method based on NARX model fitting.

We start by defining a simple criterion called \emph{mergeable}:
\begin{definition}[Mergeable Traces]\label{def:mergeable}
A set $S$ of discrete-time trace segments is \emph{mergeable} w.r.t.\ the template NARX model $\narx$ if there exists $\narxIns \in \narxInsSet$ such that $\narxIns \models S$.
\end{definition}%

Determining whether two trace segments are mergeable or not boils down to fitting multiple trace segments into an NARX model as established in \cref{subsec:narx-fitting}. Based on this criterion, one can cluster all the trace segments $\segments_{\data} = \bigcup_{j=1}^{M} \segments_{\traceDT_j}$ obtained via segmentation by finding a \emph{maximal partitioning} $\segments_{\data} = S_1 \uplus S_2 \uplus \cdots \uplus S_c$ with $c \in \PosNats$ clusters such that all segments within the same cluster $S_i$ are mergeable yet segments across different clusters are not. In contrast to trace similarity-based clustering, our method ensures that all trace segments assigned to the same cluster can be generated by the same mode dynamics (provided that the given template suffices to fit these trace segments), as guaranteed by \cref{thm:exact-fitting}.
%mergeable.

\begin{algorithm}[t]
    \caption{Trace Segmentation via NARX Model Fitting}
    % \caption{Trace Segmentation Based on NARX Model Fitting (via sliding window of size $\windowSize$)}
    \label{alg:segmentation-sliding}
    % \footnotesize
    \SetKwInput{Input}{Input}\SetKwInOut{Output}{Output}\SetNoFillComment
    \Input{A discrete-time trace $\traceDT$ of length $\traceDTLen+1$ %per \cref{eq:learning-data} 
    and a template NARX model $\narx$.}
    \Output{A set $\segments_{\traceDT}$ of trace segments of $\traceDT$ under $\narx$.}
    $\cpoints \gets \{0\}$; $l \gets 0$ \tcp*{$\mathtt{initialization;}\;l\;\mathtt{is\;left\;bound\;of\;sliding\;window}$}
    % \tcp*{initialize $\cpoints$ and $l$ (left bound of sliding window)}
    \While{$l \leq \traceDTLen + 1 - \windowSize$} {
        \If(\tcp*[f]{$\mathtt{detected\;potential\;changepoint}$}
        % \tcp*[f]{detected potential changepoint at $l + \windowSize - 1$}
        ){$\traceDT_{l, l + \windowSize}$ is not fittable by $\narx$}{
            $\cpoints \gets \cpoints \cup \{l + \windowSize - 1\}$ \tcp*{$\mathtt{update\;the\;changepoints}$}
            % \tcp*{update the changepoints}
            $l \gets l + \windowSize$ \tcp*{$\mathtt{skip\;data\;points\;in\;the\;current\;window}$}
            % \tcp*{skip data points in the current window}
        }
        \Else{
            $l \gets l + 1$ \tcp*{$\mathtt{slide\;the\;window\;to\;the\;right}$}
            % \tcp*{slide the window to the right}
        }
    }
    \tcc{$\mathtt{segment}\;\traceDT\;\mathtt{as\;per}\;\cpoints$}
    $\segments_{\traceDT} \gets \{\traceDT_{p_0, p_1}, \traceDT_{p_1, p_2}, \ldots, \traceDT_{p_{s-1}, p_s}\} \quad \text{for} \quad \cpoints = \{p_0, p_1,\ldots,p_s\}$ 
    
    % \tcp*{segment $\traceDT$ as per $\cpoints$}
    \smallskip
    \For{$\segment \in \segments$\label{alg:segmentation-sliding-check}}{
        \If{$\segment$ is not fittable by $\narx$\label{alg:segmentation-sliding-nonfittable}}{
                $\segments_{\traceDT} \gets \segments_{\traceDT} \setminus \{\segment\}$ \tcp*{$\mathtt{drop\;non\!-\!fittable\;segments\;in}\;\segments_{\traceDT}$}
                % \tcp*{drop non-fittable segments in $\segments_{\traceDT}$} 
                \label{alg:segmentation-sliding-drop}
            }
    }
    \Return $Seg_{\traceDT}$\;
\end{algorithm}

The notion of mergeable is simple; it guarantees that all %trace 
segments generated by the same %system 
mode are mergeable. However, the reverse does not hold: The set of mergeable segments may actually be generated by \emph{different} mode dynamics; see example in \ifarxiv\cref{sec:appendix-mmergeable}\else\cite[Appx.~C]{DBLP:journals/corr/abs-2507-16426}\fi. To address this, {\ourmethod} provides a less aggressive merging criterion called \emph{minimally mergeable} as detailed in \ifarxiv\cref{sec:appendix-mmergeable}\else\cite[Appx.~C]{DBLP:journals/corr/abs-2507-16426}\fi.

\subsection{Mode Characterization}\label{subsec:mode-characterization}

Given the clustered segments $\segments_{\data} = S_1 \uplus S_2 \uplus \cdots \uplus S_c$, %the process of 
mode characterization aims to construct the corresponding set of system modes $\modes_{\data} = \{q_1, q_2, \ldots, q_c\}$ by learning a NARX model $\narxIns_i$ adhering to the given template $\narx$ as the dynamics for each cluster $S_i$ in mode $q_i$ ($1\leq i \leq c$). We achieve this by applying our LLSQ-based trace fitting technique in \cref{subsec:narx-fitting}: We find $\narxIns_i \in \narxInsSet$ such that $\narxIns_i \models S_i$ by solving the linear optimization problem \cref{eqn:optimization} with \cref{eq:extended-matrices}.
% After clustering is completed, a difference equation needs to be assigned to each cluster to express the flow relation. Suppose the clustering results are \( S_1, S_2, \dots \). For each \( S_i = {t_1, t_2, \dots} \), the coefficient vector \( \mathcal{A}_i = \mathcal{A}_{t_1|t_2|\dots} \) of the difference equation can be obtained using the difference template \( \tilde{\narx} \) and the corresponding method in Equation \ref{multiple fit}, thereby forming the system of difference equations \( \narx_i = (k, X, f, \mathcal{A}_i) \). 
% Note that if the minimally mergeable criterion is chosen for %segment 
% clustering, we will instead find $\narxIns_i$ of order $\hat{k}_{S_i}$ that fits $S_i$.

The correctness of mode characterization follows directly from that %is a direct consequence of the correctness 
of the previous steps: Since our segmentation procedure ensures that every trace segment in $\segments_{\data}$ is fittable by the template $\narx$ whilst the clustering process guarantees that all the segments in \( S_i \) are %(minimally) 
mergeable, we are able to find $\narxIns_i$ %(of order $\hat{k}_{S_i}$) 
that fits $S_i$ (cf.\ \cref{thm:exact-fitting}).

\subsection{Guard Learning}\label{subsec:guard-learning}

Guard learning aims to learn a guard condition $g \subseteq \RR^{n+m}$ for each mode switching. This task can be viewed as a (binary) classification problem to predict if a transition $e$ between two modes will occur or not. Specifically, given two modes $q, q' \in \modes_{\data}$, guard learning identifies the \emph{decision boundary} that separates the region in the continuous state space $\RR^{n+m}$ where the system will transition to $q'$ from the region where it either remains in $q$ or jumps to a mode different from $q'$. Recall that we assume a \emph{deterministic} transition semantics, i.e., guards for the same (source) mode do \emph{not} overlap. %in $\RR^{n+m}$.
% \footnote{\color{red}Dainarx aims to learn \emph{deterministic} models from data, resolving guard overlaps by prioritizing the transition with the maximum SVM distance from the current state}

Considering the potential nonlinearity of guards $g \in G$ in the original hybrid automaton $\ha$, we use SVMs \cite{VapLer63,Boser92} to effectively identify the decision boundaries similarly as \cite{DBLP:journals/fac/JinAZZZ21,DBLP:conf/atva/GurungWS23}. Let $\samples = \samplesPos \uplus \samplesNeg$ be a training dataset %consisting 
of positive samples $\samplesPos$ and negative samples $\samplesNeg$. For linear decision boundaries, SVM yields the optimal hyperplane by maximizing the margin between points in $\samplesPos$ and $\samplesNeg$, thus exhibiting strong generalization capabilities; For the nonlinear case
% decision boundaries 
(where $\samplesPos$ and $\samplesNeg$ may not be linearly separable), SVM can still yield effective classification using space transformations and kernel tricks (e.g., polynomial or RBF kernels) \cite{Boser92}.

{\makeatletter
\let\par\@@par
\par\parshape0
\everypar{}
\begin{wrapfigure}{r}{0.36\linewidth}
\vspace*{-4mm}
    \centering
    \includegraphics[width=1\linewidth]{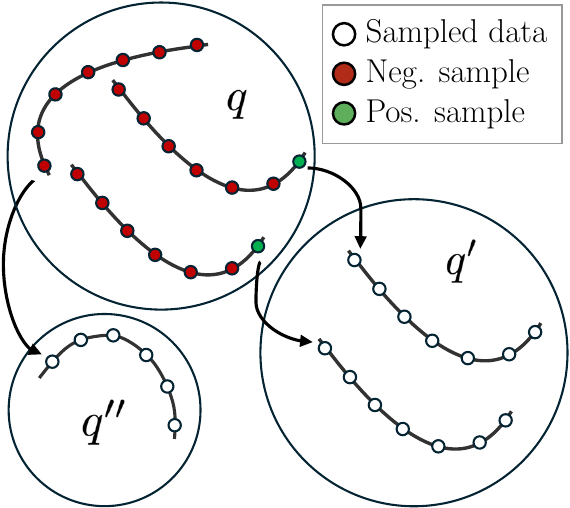}
    \vspace*{-5.6mm}
    \caption{SVM samples.}
    \label{fig:guard-learning}
\vspace*{-3.5mm}
\end{wrapfigure}
% \begin{figure}[t]
%     \centering
%     \vspace*{-5mm}
%     \includegraphics[width=0.34\linewidth]{paper/fig/svm.pdf}
%     \vspace*{-2mm}
%     \caption{Samples in $(q,q')^+, (q,q')^-$.}
%     \label{fig:guard-learning}
%     \vspace*{-4mm}
% \end{figure}

Below, we show how our guard learning problem can be reduced to SVMs by constructing the positive and negative datasets of samples. Let $\traceDT_j(\tau)$ be the $\tau$-th timestamp-free data point $\vec{v}_\tau$ on trace $\traceDT_j$; let $\targetMode_{j}(\tau)$ be the mode $q_{i} \in \modes_{\data}$ such that the trace segment containing $\traceDT_j(\tau)$ is clustered into $S_i$.\footnote{$\targetMode_{j}(\tau)$ does not exist if the %trace
segment containing $\traceDT_j(\tau)$ is ruled %screened 
out in %during
segmentation.} Then, for 
\par}
% \begin{align*}
%     \left(q,q'\right)^+ &\eeq \bigcup\nolimits^M_{j=1}\left\{ \traceDT_j(\tau) \mid \targetMode_{j}(\tau) = q \ \, \text{and}\ \, \targetMode_{j}(\tau+1) = q' \right\}~,\qquad\qquad\qquad\qquad\qquad\qquad\qquad\ \\
%     \left(q,q'\right)^- &\eeq \bigcup\nolimits^M_{j=1}\left\{ \traceDT_j(\tau) \mid \targetMode_{j}(\tau) = q \ \, \text{and}\ \, \targetMode_{j}(\tau+1) \neq q' \right\}~.
% \end{align*}%
\noindent
any pair of modes $q, q' \in \modes_{\data}$, the corresponding positive dataset $(q,q')^+$ and negative dataset $(q,q')^-$ are constructed:
\begin{align*}
    \left(q,q'\right)^+ &\eeq \bigcup\nolimits^M_{j=1}\left\{ \traceDT_j(\tau) \mid \targetMode_{j}(\tau) = q \ \, \text{and}\ \, \targetMode_{j}(\tau+1) = q' \right\}~, \\
    \left(q,q'\right)^- &\eeq \bigcup\nolimits^M_{j=1}\left\{ \traceDT_j(\tau) \mid \targetMode_{j}(\tau) = q \ \, \text{and}\ \, \targetMode_{j}(\tau+1) \neq q' \right\}~.
\end{align*}%
% {
% \setlength{\abovedisplayskip}{8pt}
% \setlength{\belowdisplayskip}{8pt}
% \[
% \left(q,q'\right)^+ \eeq \bigcup\nolimits^M_{j=1}\left\{ \traceDT_j(\tau) \mid \targetMode_{j}(\tau) = q \ \, \text{and}\ \, \targetMode_{j}(\tau+1) = q' \right\}~,
% \]
% \[
% \left(q,q'\right)^- \eeq \bigcup\nolimits^M_{j=1}\left\{ \traceDT_j(\tau) \mid \targetMode_{j}(\tau) = q \ \, \text{and}\ \, \targetMode_{j}(\tau+1) \neq q' \right\}~.
% \]
% }%
\noindent
Note that in $(q,q')^+$, $\tau+1$ is in fact a changepoint identified on trace $\traceDT_j$; moreover, we have $(q,q')^+ = \emptyset$ if our learning data $\data$ does not reflect a transition from $q$ to $q'$. \cref{fig:guard-learning} illustrates the intuition of constructing  $(q,q')^+$ and $(q,q')^-$.

% \noindent {\textbf{SVM For Guard:}} For mode ${\bf q}_i,{\bf q}_j$, where $i, j\in[1,M]$ and $M$ is the number of modes. SVM model $SM_{ij}$ takes the current states ${\bf x}_1,\cdots,{\bf x}_n$ as input and predicts whether the transition occurs, where $n$ is the number of system variables. As we need to learn an SVM model for each pair of modes that have a transition relationship, the maximum number of models learned is $M{\times}M$.

% \noindent {\textbf{SVM Training:}} The results of the previous steps identify which modes have transition relationships between them and provide a training set for SVM models. The following feature vector and label are used:

% \begin{equation}
%     \label{SVM-train}
%     \begin{split}
%     \bf{feature} &= [x[t],\forall{i} \in [n]] \\ 
%     l&abel = \pm 1
%     \end{split}
% \end{equation}
% where 
% \begin{itemize}
%     \item $x[t],\forall{i} \in [n]$ gives the value of all system variables at time $t$,
%     \item $1$ indicates a mode transition, while $-1$ represents no occurrence of transition.
% \end{itemize}
% To be more specific, the detected changepoints from $q_i$ to  $q_j$ are labeled as positive for $SM_{ij}$ and negative for $SM_{ik}$, where $k \in [1, M]$ and $k \neq i, j$. For points that are not a changepoint, they are all labeled as negative for $SM_{i*}$, if current mode is $q_i$.

\subsection{Reset Learning}

The final step aims to fill the hole in every transition $e  = (q, g, \cdot, q') \in \trans$ by learning a \emph{linear} reset function $r\colon\RR^{k \times n + m} \to \RR^{k \times n}$ such that the output state vectors $\vec{x}, \vec{x}^{(1)}, \ldots, \vec{x}^{(k-1)}$ are updated to $r(\vec{x}, \vec{x}^{(1)}$, $\ldots, \vec{x}^{(k-1)}, \vec{u})$ upon taking this transition (triggered by the activation of $g$, i.e., $(\vec{x},\vec{u}) \in g$).

For the simple case of first-order dynamics with $k = 1$, learning the reset function can be done via \emph{linear regression} to capture the relation between the data points before and after the transition, regardless of the type of dynamics; %(linear or nonlinear); 
see, e.g., LearnHA \cite{DBLP:conf/atva/GurungWS23}.
%for more details.

However, for the case of high-order dynamics with $k > 1$, the same method based on linear regression does not apply because the values of higher-order derivatives ($\vec{x}^{(1)}, \ldots, \vec{x}^{(k-1)}$) are not present in our learning data $\data$. Therefore, we employ a different form of reset function $r\colon \RR^{k \times n + m} \to \narxInsSet^k\,$ as 
\begin{equation}
\label{eq:reset-k-narx}
    \left(\vec{x}[\tau],  \ldots, \vec{x}[\tau - k + 1], \vec{u}[\tau]\right) \,\mapsto\, \left\{\narxIns_1, \ldots, \narxIns_k\right\}~,
\end{equation}
such that each $\narxIns_i\colon (\vec{x}[\tau - 1 + i], \ldots, \vec{x}[\tau - k + i], \vec{u}[\tau + i]) \mapsto \vec{x}[\tau + i]$ can be used to generate $\vec{x}[\tau + i]$ for the new mode. Next, we show how to construct the $k$ NARX models in \cref{eq:reset-k-narx} as the reset of transition $e  = (q, g, \cdot, q')$ via NARX model fitting. For any $1 \leq i \leq k$,
\begin{align*}
    \narxIns_i \mmodels \bigcup\nolimits^M_{j=1}\left\{\traceDT_{l, l + k + 1} \subseteq \traceDT_j \mid \traceDT_j(l + k - i) \in {(q, q')}^+\right\}~,
\end{align*}%
where $\traceDT \subseteq \traceDT'$ means that $\traceDT$ is a segment of $\traceDT'$; that is, we fit $\narxIns_i \in \narxInsSet$ using the data points on the $i$-th segment (of length $k+1$) enclosing the mode-switching point.

For hybrid systems with linear dynamics, we \emph{conjecture} that -- if the sampling rate of our learning data $\data$ is beyond the Nyquist frequency \cite{diniz2010digital} -- the above reset learning technique based on NARX model fitting is guaranteed to identify the correct resets (subject to %the learning data 
$\data$). The rationale behind this conjecture is the Nyquist-Shannon sampling theorem \cite{diniz2010digital} in signal processing and the Picard–Lindel{\"o}f theorem (aka, the Cauchy–Lipschitz theorem) \cite{arnold1992ordinary} for initial value problems -- these two results \emph{may} be used to establish that one can perfectly reconstruct the original continuous-time signal adhering to linear ODEs from (sufficiently dense) data samples.

For nonlinear dynamics, due to the complex nature of the induced signal, we do not foresee a similar conjecture as for the linear case. However, we still observe effective approximations of the reset functions in practice, as demonstrated in \cref{sec:overview,subsec:experiments-nonlinear}.

\paragraph*{\bf Time Complexity} The total \emph{worst-case} time complexity of {\ourmethod} is $\bigO(\abs{\data}^3 \cdot n + \abs{\data}^2 \cdot d^2 \cdot n)$ (see details in \ifarxiv \cref{sec:appendix-complexity}\else\cite[Appx.~D]{DBLP:journals/corr/abs-2507-16426}\fi), i.e., \emph{polynomial} in the size of the learning data $\data$ and the size of the target system $d = (\alpha + kn + m + 1)$. In practice, however, the complexity is much lower (bounded by $\bigO(\abs{\data} \cdot d^2 \cdot n \cdot (\windowSize+\abs{\segments_\data})$).
\section{Extending {\ourmethod} for Noisy Data}\label{sec:noisy}

Motivated by \cite{kochdumper2025robust}, we now unleash {\ourmethod} for its general application to the \emph{robust inference of hybrid automata from noisy data}. The keys to such extension are \emph{resampling} for segmentation (\cref{sec:resampling}) and \emph{total least squares} (TLS) for mode characterization (\cref{sec:tls}).

\subsection{Models under Noise}

We extend the ODE-modeled dynamics \cref{eq:odes} as a simplified system of high-order \emph{stochastic differential equations} (SDEs) of the form
\begin{align}\label{eq:sdes}
\vec{x}^{(k)} \eeq \,F_q \left(\vec{x}, \vec{x}^{(1)}, \ldots, \vec{x}^{(k-1)},\vec{u}\right) \pplus \sqrt{p} \cdot \vec{\omega}^{(1)}~,
\end{align}%
where $\vec{\omega} \in \Reals^n$ denotes a standard Wiener process (Brownian motion), $p \in 
\Reals$ is the scalar noise spectral density, and $\vec{\omega}^{(1)}$\! models the \emph{process noise} in the form of Gaussian white noise.

Similarly, for measuring the state $\vec{x}$ in \cref{eq:sdes}, we have
\begin{equation*}
\vec{y} \eeq \vec{x} \pplus \sqrt{m} \cdot \vec{\nu}^{(1)}~,
\end{equation*}%
where $\vec{\nu}^{(1)}$\! denotes the \emph{measurement noise} and $m \in 
\Reals$ the corresponding spectral density. We assume that, for each state variable, the process and the measurement noise are independent and identically distributed (i.i.d.), mutually independent, and both Gaussian. Moreover, by sampling with period $\Delta t$, the variance of process (measurement, resp.) noise is $\sigma_{\mathit{proc}}^2 = p \cdot \Delta t$ ($\sigma_{\mathit{proc}}^2 = m \cdot \Delta t$, resp.).

In order to characterize the above flow dynamics under noise, we extend our NARX model \cref{eq:narx-standard} in a similar vein as 
\begin{equation}\label{eq:noisy_narx_dynamic}
    \vec{x}[\tau] = F_q\left(\vec{x}[\tau - 1], \vec{x}[\tau - 2],\ldots, \vec{x}[\tau - k], \vec{u}[\tau]\right)\! \pplus \vec{\omega}[\tau] %\quad \text{for}\ \tau \geq k
\end{equation}%
for $\tau \geq k$, where $\vec{\omega}[\tau] \stackrel{\mathrm{i.i.d.}}{\sim} \mathcal{N}_n(0, \sigma_{\mathit{proc}}^2 \cdot I_n)$ is the Gaussian-distributed process noise ($I_n$ is the identity matrix of dimension $n$). 

The measurement of $\vec{x}[\tau]$ in \cref{eq:noisy_narx_dynamic} is
\begin{equation}\label{eq:noisy_narx_observe}
\vec{y}[\tau] \eeq \vec{x}[\tau] \pplus \vec{\nu}[\tau] \quad \text{with} \quad \vec{\nu}[\tau] \stackrel{\mathrm{i.i.d.}}{\sim} \mathcal{N}_n(0, \sigma_{\mathit{proc}}^2 \cdot I_n)~.
\end{equation}

Note that $\vec{\omega}[\tau]$ in \cref{eq:noisy_narx_dynamic} can be regarded as an unobservable white-noise input, while $\vec{\nu}[\tau]$ in \cref{eq:noisy_narx_observe} obscures the true state values.

\subsection{Resampling for Segmentation}\label{sec:resampling}

In the presence of noise, it becomes difficult, if not impossible, to distinguish model-mismatch errors ($E_S$ in \cref{thm:exact-fitting}) from noise-induced errors, thereby undermining the steps based on NARX model fitting. This issue is especially prominent for the segmentation step as it uses only a few data points within the sliding window.
% and the segmentation step that locates switching points from fit residuals may fail. 
To address this issue, we employ \emph{resampling} to amplify errors induced by model mismatch. Specifically, given a resampling stepsize $d>1$, we use the following \emph{resampled trajectory} $\xi|_{d}$ for any $\xi \in \data$:
\begin{equation*}
\xi|_{d} \ddefeq
\left\langle
(t_0, \vec{v}_0),
(t_0 + d\cdot \Delta t, \vec{v}_{d}),\,
\ldots,\,
\big(t_0 + \ell_{d}\cdot \Delta t, \vec{v}_{\ell_{d}}\big)
\right\rangle~,
\end{equation*}%
to identify changepoints in segmentation, where $\ell_{d} = \left\lfloor \nicefrac{\ell}{d} \right\rfloor \cdot d$.

% The resampled trajectory has a larger step size, which enlarges mismatch-induced errors and thereby makes them more distinguishable from noise. We first resample the trajectory and then apply the segmentation step to compute the changepoints.

\subsection{TLS for Mode Characterization}\label{sec:tls}

Now, consider the impact of noise on the LLSQ used in model fitting: For the process noise, substituting \cref{eq:noisy_narx_dynamic} into \cref{eqn:optimization}, when fittable, the remaining optimization objective becomes $\vec{\omega}[\tau]$ whose expected value is $0$. This indicates that, with only process noise, the ordinary least squares remain a good unbiased estimator (with large sample size) -- i.e., process noise does not bias our fit; For the measurement noise, however, substituting \cref{eq:noisy_narx_observe} into \cref{eqn:optimization} yields a biased fit.
%, because the observed values are not the true $\vec{x}[\tau]$, the objective can no longer be simplified, and LS becomes biased.

To remedy this, we replace \cref{eqn:optimization} with \emph{total least squares} (TLS) \cite{golub1980analysis}:
\begin{mini*}|l|[2]
    {\Lambda_{i,:}, \Delta O_{i,:}, \Delta D_i}
    {\big\lVert \Delta O_{i,:} \big\rVert_{F} \pplus \big\lVert \Delta D_i \big\rVert_{F} \quad \quad \quad \quad \quad \quad \quad}
    {\label{eqn:optimization_tls}}
    {}
    \addConstraint{\normDisplay{\left( O_{i,:} + \Delta O_{i,:} \right) - \Lambda_{i,:} \cdot \left( D_i + \Delta D_i \right)} \eeq 0}
\end{mini*}%
where $\lVert \cdot \rVert_F$ is the Frobenius-norm. Intuitively, TLS seeks the \emph{smallest offsets} $\Delta O_{i,:} \!\in\! \mathbb{R}^{1 \times (\ell-k+1)}, \Delta D_i \!\in\! \mathbb{R}^{(\alpha + k n + m + 1) \times (\ell - k + 1)}$ as compensation for their noisy counterparts $O_{i,:}, D_i$ to retrieve fittability.

% $\left( O_{i,:} + \Delta O_{i,:} \right) - \Lambda_{i,:} \times \left( D_i + \Delta D_i \right) = 0$, while minimizing $\left\|\begin{bmatrix}\Delta O_{i,:} \! & \! \Delta D_i\end{bmatrix}\right\|_F$ (equivalently, $\left\| \Delta O_{i,:} \right\|_2 + \left\| \Delta D_i \right\|_F$). where $\left \| \cdot \right \|_F$ is the Frobenius-norm. The optimization problem can be formalized as

% For linear NARX models, $\Delta O_{i,:}$ and $\Delta D_i$ correspond to measurement noise and process noise, respectively. 
Like LLSQ, TLS admits efficient off-the-shelf solvers \cite{golub1980analysis}. Note, however, that TLS is not directly applicable to nonlinear ARX models (which again induce a biased fit). Thus, for the nonlinear case, we retain LLSQ. Moreover, TLS introduces additional degrees of freedom ($\Delta O_{i,:}, \Delta D_i$) and thus requires more data for a good fit. Hence, we use TLS for mode characterization other than segmentation.

% !TEX root = main.tex

\begin{table*}[t]
  \centering
  \setlength{\tabcolsep}{2pt} 
  %\scriptsize 
  % \begin{threeparttable}
  \caption{Results on inferring HA with linear (upper part) and nonlinear (lower part) dynamics/guards. $\abs{\modes}$, $\abs{\varsO}$, and $k$ represent the number of modes, the number of output variables, and the highest order of ODEs, respectively; $\hdt$ is the Hausdorff distance (in time) of the sequence of identified changepoints with the sequence of genuine mode-switching points; 
  % (irrelevant for RIHAND as it conducts no changepoint detection); 
  $\diffmax$ ($\diffavg$, resp.) represents the maximal (average, resp.) point-wise difference between the traces of the inferred system and the original traces (ground truth). Total Time is the total execution time of the tools. For linear HA, \benchmark{tanks} contains 1 input variable; \benchmark{ball} involves linear resets; \benchmark{underdamped-c} contains both linear resets and an input variable; For nonlinear HA, \benchmark{lotkaVolterra}, \benchmark{spacecraft}, and \benchmark{duffing} feature nonlinear guards; \benchmark{oscillator} contains trigonometric dynamics and guards; \benchmark{duffing} has linear resets and inputs. \enquote{--} marks inference failure.}
  %indicates that a tool does not suffice for the inference task of the corresponding benchmark.
  \label{tab:merged_table}
  \resizebox{\textwidth}{!}{
  \begin{tabular}{lcccrrrrrrrrrrrrrrrr}
   \toprule
    \multicolumn{4}{c}{Benchmark Details}  & \multicolumn{4}{c}{$\hdt$ (seconds)%$^\ddagger$
    } & \multicolumn{4}{c}{$\diffmax$} & \multicolumn{4}{c}{$\diffavg$} & \multicolumn{4}{c}{Total Time (seconds)} \\
    \cmidrule(lr){1-4} \cmidrule(lr){5-8} \cmidrule(lr){9-12} \cmidrule(lr){13-16}\cmidrule(lr){17-20}
    Name & $\abs{\modes}$ & $\abs{\varsO}$ & $k$ & LearnHA & FaMoS & RIHAND &\textbf{\ourmethod} & LearnHA & FaMoS & RIHAND& \textbf{\ourmethod} & LearnHA & FaMoS &RIHAND& \textbf{\ourmethod} & LearnHA & FaMoS & RIHAND& \textbf{\ourmethod} \\
    \midrule
    \benchmark{buck\_converter}~\cite{DBLP:conf/hybrid/PlambeckBHF24} & 3 & 2 & 1 & 0.0001 & 0.0004 &0.0194 & \textbf{0.0000} & 0.1674 & 0.2065 &1.0571& \textbf{0.0000}& 0.0046 & 0.0142 & 0.0670&\textbf{0.0000}& 28.70 & 4.43 & 31.07 &\textbf{2.12} \\
    \benchmark{complex\_tank}~\cite{DBLP:conf/hybrid/PlambeckBHF24} & 8 & 3 &1& 3.6350 & 1.6350 & 2.6700 & \textbf{0.0950} & 0.3483 & 0.2411 &0.3060& \textbf{0.0451}  & 0.0655 & 0.0151 &0.0539&\textbf{0.0028} & 40.50 &  19.00 & 161.03&\textbf{5.73} \\
    \benchmark{multi\_room\_heating}~\cite{DBLP:conf/hybrid/PlambeckBHF24} & 4 & 3 &1& 2.9750 & 0.0600 &0.8400& \textbf{0.0350} & 0.2531 & 0.0094 &0.2432& \textbf{0.0078} & 0.0590 & 0.0013 &0.0641& \textbf{0.0008} &  23.30 & 8.12 & 63.29 & \textbf{2.96} \\
    \benchmark{simple\_heating\_syst}~\cite{DBLP:conf/hybrid/PlambeckBHF24} & 2 & 1 &1& 0.0400 & 0.4400 & 0.8600 &\textbf{0.0200} & 0.0202 & 0.2126 &0.1828&\textbf{0.0102}  & 0.0037 & 0.0423 &0.0600& \textbf{0.0007} & 13.40 & 8.39 & 15.23 & \textbf{0.78} \\
    \benchmark{three\_state\_HA}~\cite{DBLP:conf/hybrid/PlambeckBHF24} & 3 & 1 & 2 & -- & 0.5700 & 9.2300& \textbf{0.0000} & -- & 0.5388&1.7687 & \textbf{0.0000} & -- & 0.0178 &0.1448& \textbf{0.0000} & -- & 1.45 &37.40 &  \textbf{0.92} \\
    \benchmark{two\_state\_HA}~\cite{DBLP:conf/hybrid/PlambeckBHF24} & 2 & 1 & 2 & -- & 0.3400 &3.4800& \textbf{0.0000} & -- & 0.1284&0.1014 & \textbf{0.0000} & -- & 0.0132 &0.0125& \textbf{0.0000} & -- & 1.72 &25.14 & \textbf{0.80} \\
    
    \benchmark{variable\_heating\_syst}~\cite{DBLP:conf/hybrid/PlambeckBHF24} & 3 & 2 & 1 & 0.1100 & 0.0700 &6.9200 & \textbf{0.0300}&  0.0581 & 0.0272 & 302.8669&\textbf{0.0200}  & 0.0082 & 0.0012 & 20.4129&\textbf{0.0005} & 11.30 & 4.23 &31.75 & \textbf{1.79} \\
    
    \benchmark{cell}~\cite{DBLP:conf/atva/GurungWS23} & 4 & 1 & 1 & \textbf{0.0100} & 29.2800 & 95.1800 &\textbf{0.0100}& \textbf{0.0176} & 1.6144 &3.3968 &\textbf{0.0176} &\textbf{0.0001}& 0.1494 & 0.5590&0.0002 & 65.20 & 40.42 &78.47& \textbf{7.47} \\
    
    \benchmark{oci}~\cite{DBLP:conf/atva/GurungWS23}& 2 & 2 & 1& 0.0500 & -- &1.3900 & \textbf{0.0000} & 0.2002 & -- & 0.8375&\textbf{0.0000} & 0.0259 & -- & 0.1510&\textbf{0.0000}& 8.46 & -- & 23.81 & \textbf{1.41} \\
    
    \benchmark{tanks}~\cite{DBLP:conf/atva/GurungWS23} & 4 & 2 & 1 & 13.9100 & -- & 17.9200 & \textbf{0.0100} & 1.1077 & -- &1.3229& \textbf{0.0177} & 0.2589 & -- & 0.2059&\textbf{0.0007} & 20.79 & -- & 19.33 & \textbf{3.82} \\
    
    \benchmark{ball}~\cite{DBLP:conf/atva/GurungWS23}& 1 & 2 & 1 & \textbf{0.0000} & -- &\textbf{0.0000} & \textbf{0.0000} & \textbf{0.0000} & -- &1.8911& \textbf{0.0000} & \textbf{0.0000} & -- & 0.0759&\textbf{0.0000}  & 3.38 & -- & 9.95 &\textbf{0.72} \\
    
    \benchmark{dc\_motor} & 2 & 2 & 4 & -- & -- & -- & \textbf{0.0000} & -- &-- & -- & \textbf{0.0000}& -- & -- & --&\textbf{0.0000} & -- & -- & --& \textbf{4.08} \\
    
    \benchmark{simple\_linear} & 2 & 2 & 1 & 7.9600 & 0.0600 & 1.8400 & \textbf{0.0000} & 0.9999 & 0.1107 & 0.5013 &\textbf{0.0000} & 0.1448 & 0.0071 & 0.0407&\textbf{0.0000}& 12.79 & 3.80 & 25.87 & \textbf{1.97} \\
    
    \benchmark{jumper}~\cite{DBLP:conf/rss/Borquez0C0B24} & 2 & 4 & 1 & 0.4500 & -- & 3.4800& \textbf{0.0000} & 1.8182 & -- & 7.4120&\textbf{0.0000} & 0.0899 & -- &1.2111&\textbf{0.0000} & 3.35 & -- &13.90 & \textbf{2.38} \\

    \benchmark{loop\_syst} & 4 & 2 & 2 & -- & 4.9100 & 6.6100 & \textbf{0.0000} & -- & 1.8929 & 0.8548 &\textbf{0.0000} & -- & 0.2470 &0.1084&\textbf{0.0000} & -- &  3.30 & 77.34 & \textbf{3.10} \\
    
    \benchmark{two\_tank}& 2 & 2 & 1 &\textbf{0.0000} & 8.6400 & 10.4800 & \textbf{0.0000}& \textbf{0.0000} & 1.5273 &0.0286 &\textbf{0.0000}& \textbf{0.0000} & 0.2324 &0.0078& \textbf{0.0000} & 7.42 &  \textbf{2.94} & 24.16 &4.61 \\
    
    \benchmark{underdamped}& 2 & 2 & 2 & -- & -- &15.0200& \textbf{0.0000} & -- & -- &4.0053 &\textbf{0.0000}& -- & -- & 0.9305& \textbf{0.0000} & -- & -- & 77.98& \textbf{2.25} \\
    \benchmark{underdamped-c} & 4 & 2 & 2 & -- & -- & -- & \textbf{0.0100}& -- & -- & --&\textbf{0.0080} & -- & --&-- &\textbf{0.0004}  & -- &  --& -- &\textbf{4.52}  \\ 
    
    \benchmark{super\_complex\_tank} & 12 & 3 & 1 & 3.2150 & 1.4700 & 12.4800& \textbf{0.0950} & 0.3723 & 0.2412 & 3.5300& \textbf{0.0454} & 0.0784 & 0.0127 & 0.4216 & \textbf{0.0035}  & 41.39 & 18.66 & 127.68 & \textbf{5.43} \\
        
    \midrule
    \benchmark{lander}~\cite{DBLP:conf/fm/ZhaoYZGZC14} &2 &4&1 & \multirow{8}{*}{--} & \multirow{8}{*}{--} & 0.3900&\textbf{0.0100} &  \multirow{8}{*}{--}& \multirow{8}{*}{--} &0.0035 & \textbf{0.0024 }& \multirow{8}{*}{--} & \multirow{8}{*}{--} &0.0003 & \textbf{0.0000} & \multirow{8}{*}{--} & \multirow{8}{*}{--} & 26.53& \textbf{2.58} \\
    
    \benchmark{lotkaVolterra}~\cite{DBLP:conf/arch/GerettiSABCFILM23} &2&2& 1&& & 3.1400 &\textbf{0.0200} & & &0.5152 & \textbf{0.0047} & &&0.1146  & \textbf{0.0006} & &&  32.49&\textbf{0.76} \\
    
    \benchmark{simple\_non\_linear} &2&1&1& &&8.8600& \textbf{0.0060} & && 0.8973 & \textbf{0.0367} & & &0.1897& \textbf{0.0020} & && 53.71&\textbf{5.70} \\
    
    \benchmark{simple\_non\_poly} &2&1&1& && 43.6000&\textbf{0.0000} & && 0.5570& \textbf{0.0000} & && 0.1701& \textbf{0.0000} & &&78.78& \textbf{3.96} \\
    
    \benchmark{oscillator}~\cite{DBLP:conf/adhs/RamdaniN09} &2&2&1& && 0.3280& \textbf{0.0000} & && 0.0162& \textbf{0.0000} && & 0.0006& \textbf{0.0000} && & 28.87&\textbf{1.13} \\
    
    \benchmark{spacecraft}~\cite{DBLP:conf/cpsweek/ChanM17} &2&4&1&& & 2.0800&\textbf{0.0000} & && 0.1959&\textbf{0.0000} & && 0.0190 & \textbf{0.0000} & &&38.96& \textbf{2.01} \\
    
    \benchmark{sys\_bio}~\cite{DBLP:journals/iandc/WangCXZK22} &2&9&1& && 0.9730& \textbf{0.0120} & & & 0.5244 & \textbf{0.0960} & && 0.0229& \textbf{0.0081} & &&239.86& \textbf{27.40} \\
    
    \benchmark{duffing} &2&1&2& && 1.5030&\textbf{0.0010} & && 1.8358& \textbf{0.0003} & && 0.3042& \textbf{0.0000} & && 150.93&\textbf{7.34} \\
    \bottomrule
  \end{tabular}
  % \begin{tablenotes}[flushleft]
  %     \small
  %     \item[$\ddagger$] $\hdt$ does not apply to RIHAND as it conducts no mode-switching detection.
  %   \end{tablenotes}
  % \end{threeparttable}
  }
\end{table*}

\section{Implementation and Experimental Results}\label{sec:experiments}

We have implemented {\ourmethod} as a prototype tool %(of the same name) 
in Python 3.9.\footnote{\ifanonymous Available at~\url{https://anonymous.4open.science/r/Dainarx-1CEC}. \else Available at~\faGithub~\url{https://github.com/FICTION-ZJU/Dainarx}.\fi} By interfacing with NumPy \cite{DBLP:journals/nature/HarrisMWGVCWTBS20} (for LLSQ and matrix operations) as well as scikit-learn \cite{DBLP:journals/jmlr/PedregosaVGMTGBPWDVPCBPD11} (for SVM classifications), {\ourmethod} infers a potentially nonlinear hybrid automaton with high-order NARX-modeled dynamics (under template $\narx$) from possibly noisy input-output discrete-time traces $\data$. This section reports the evaluation results of {\ourmethod} against state-of-the-art tools on a collection of benchmarks. All experiments are conducted on an 8-core Apple M2 processor with 16GB RAM.
The experiments are designed to evaluate {\ourmethod} in terms of \emph{effectiveness},  \emph{efficiency}, and \emph{robustness}.

\subsection{Experimental Setup}

\paragraph{\ourmethod Configurations}
\ourmethod supports user-configured NARX templates (see \ifarxiv\cref{sec:appendix-narx-templates}\else\cite[Appx.~E]{DBLP:journals/corr/abs-2507-16426}\fi), SVM parameters, and other experimental settings through a dedicated JSON file for each benchmark.\footnote{Detailed configurations and default parameter values are available in the artifact.}
% \footnote{The default parameter values and the configurations used in our experiments are available in the artifact.} 
We note that, in contrast to threshold-based approaches, {\ourmethod} does \emph{not} require user-supplied, fine-tuned parameter valuations for both segmentation and clustering in the noise-free setting. Instead, we employ a built-in, self-adaptive threshold $\eta = 10^{-6} \times \Delta t \times |\vec{v}|_{\max}$ to check NARX fittability against potential numerical errors, where $|\vec{v}|_{\max}$ is the maximum magnitude of the trace data. For the noisy scenario, the performance of \ourmethod relies on the adjustment of hyperparameters, e.g., the resampling stepsize $d$ as explained in \cref{sec:resampling}. In our experiments, the floating-point tolerance is set to $1\times 10^{-7}$. We use the RKF45 method (the Runge–Kutta–Fehlberg method of order 4 with an error estimator of order 5 \cite{arnold1992ordinary}) to obtain discrete-time traces %of ODEs
with fixed sampling periods; For each benchmark, we sample 15 traces -- 9 used for training (learning) the automaton and 6 used for testing the automaton.

% \paragraph{Baselines}
% We compare \ourmethod against state-of-the-art tool for hybrid system identification: FaMoS~\cite{DBLP:conf/hybrid/PlambeckBHF24} and LearnHA~\cite{DBLP:conf/atva/GurungWS23}, both of which adopt the common break down of the hybrid system learning problem into four steps as well. HAutLearn \cite{DBLP:journals/tcps/YangBKJ22} is another outstanding framework for hybrid system identification. However, as FaMoS demonstrates clear advantages over HAutLearn, we no longer conduct a comparison with HAutLearn in our experiments.

\paragraph{Baselines}
For the noise-free setting, we compare \ourmethod against three state-of-the-art implementations (with fine-tuned parameters) for hybrid system identification: FaMoS \cite{DBLP:conf/hybrid/PlambeckBHF24}, LearnHA~\cite{DBLP:conf/atva/GurungWS23}, and RIHAND~\cite{kochdumper2025robust}. FaMoS is dedicated to learning hybrid automata with high-order dynamics captured by \emph{linear} ARX models, whilst LearnHA aims to infer automata with \emph{first-order} polynomial ODEs. Both frameworks align with the methodology of decomposing the inference problem into a similar sequence of subproblems. While HAutLearn \cite{DBLP:journals/tcps/YangBKJ22} is another notable method in this domain, empirical findings in \cite{DBLP:conf/hybrid/PlambeckBHF24} demonstrate FaMoS' distinct advantages over HAutLearn in terms of performance and generalizability. For the noisy setting, we compare against RIHAND -- an approach tailored for the robust identification of hybrid automata from noisy data.
%Given this demonstrated superiority, we omit HAutLearn from our experimental comparisons. % to focus on the most competitive contemporary baselines.

% \paragraph{Benchmarks}
% We have collected a total of 25 examples\ajcomment{\atmbh\atyhz If it is possible, please add a reference/citation to each example in the following tables like what we do in \cref{tab:comparison}.} of linear and nonlinear hybrid systems. Among these, 17 are linear systems: 7 from the FaMoS benchmarks, 4 from the LearnHA benchmarks and the remaining 6 include a simple synthetic example, a Two Tanks System, Underdamped Systems with two and four modes respectively, a PID-controlled DC Motor Position System, and a hybrid system of a jumping robotic dog \cite{DBLP:conf/rss/Borquez0C0B24}. Nonlinear examples involve ODEs with dynamic behaviors characterized by polynomials, trigonometric functions, and rational functions, with guards also incorporating polynomials or trigonometric functions. The number of modes in our benchmarks typically ranges from 2 to 4, with a maximum of 8 modes in a Complex Tank System. Regarding the number of variables, most systems involve between 2 and 4 variables, while the system with the highest number of variables has 9 variables. Most of these examples are first- or second-order systems; however, we have also included a fourth-order system to demonstrate the performance of \ourmethod on higher-order systems. 

\paragraph{Benchmarks}
For the noise-free setting, we collect 27 hybrid automata from the literature. Of these, 19 are linear systems: 7 from FaMoS \cite{DBLP:conf/hybrid/PlambeckBHF24},
%\footnote{Although \benchmark{multi\_room\_heating} and \benchmark{three\_state\_ha} are collected from FaMoS \cite{DBLP:conf/atva/GurungWS23}, FaMoS did not run these cases.}, 
4 from LearnHA \cite{DBLP:conf/atva/GurungWS23}, 8 additional curated instances: a synthetic minimal system, a two-tank system, (2- and 4-mode variants of) underdamped systems, a closed-loop control system, a PID-controlled DC motor position system, a jumping robotic dog~\cite{DBLP:conf/rss/Borquez0C0B24}, and a complex tank system with 12 modes. The nonlinear subset consists of 8 hybrid systems adapted from the literature with polynomial, rational, or trigonometric dynamics as well as guard conditions involving polynomial or trigonometric constraints. The benchmarks exhibit 2–4 modes in most cases, extending to a maximum of 12 modes (\benchmark{super\_complex\_tank}); The number of output variables typically ranges from 2 to 4, with one outlier featuring 9 output variables (\benchmark{sys\_bio}). While most examples have 1st- or 2nd-order dynamics, we include a 4th-order system to evaluate \ourmethod's scalability to higher-order regimes (\benchmark{dc\_motor}).
For the noisy setting, we collect 14 benchmarks: 6 from RIHAND~\cite{kochdumper2025robust} with the same noise distributions and 8 from the noiseless benchmarks with Gaussian noise injected. \footnote{We do not reuse traces from RIHAND as they feature non-uniform sampling and regularizing them would introduce undesired interpolation noise.}
%\footnote{We excluded their traces due to methodological concerns: FaMoS employs differencing rather than Runge-Kutta methods, which we consider inaccurate, while RIHAND's traces have nonuniform time intervals that require linear interpolation for regularization, introducing undesired interpolation noise.}

\subsection{Effectiveness of {\ourmethod}}
Below, we first evaluate \ourmethod against LearnHA, FaMoS, and RIHAND -- on both linear (\cref{subsubsec:experiments-linear}) and nonlinear (\cref{subsec:experiments-nonlinear}) hybrid systems -- in terms of the \emph{accuracy of mode-switching detection}, \emph{trace fidelity}, and \emph{applicability}. Empirical results demonstrate \ourmethod's overall superiority in these metrics. Then, we showcase the rationale for the advantage of {\ourmethod} over derivative-based segmentation and trace-similarity-based clustering.

\subsubsection{Inference of Linear Hybrid Systems}\label{subsubsec:experiments-linear}

\begin{figure}[t]
    \centering
    \begin{minipage}[b]{0.47\linewidth}
        \centering
        \includegraphics[width=\textwidth]{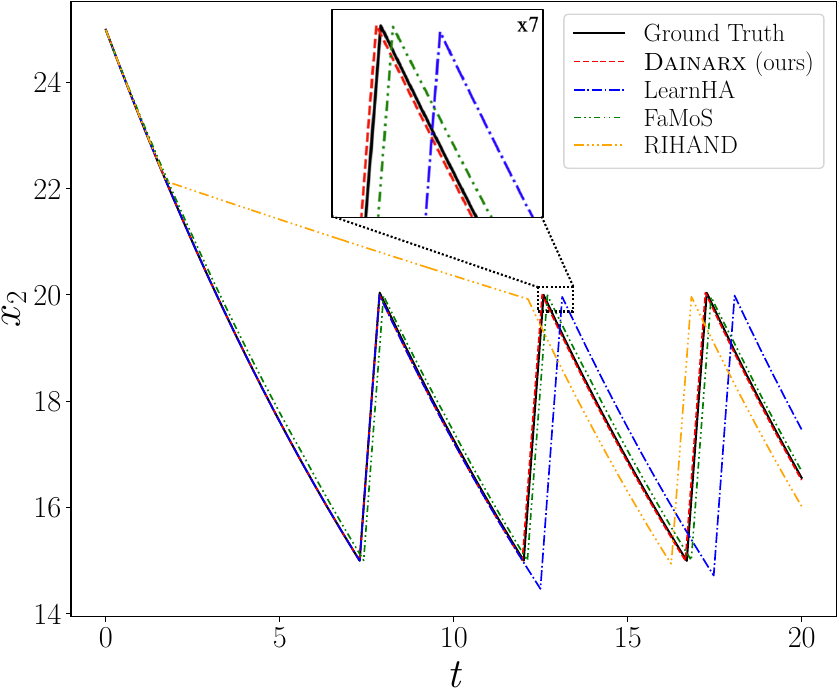}
        \caption{Traces for \benchmark{complex\_tank}.}
        \label{fig:complex_tank}
    \end{minipage}%
    \hfill
    \begin{minipage}[b]{0.506\linewidth}
        \centering
        \begin{tikzpicture}[scale=1, transform shape]
            \draw (0, 0.) node[inner sep=0] {\includegraphics[width=\textwidth]{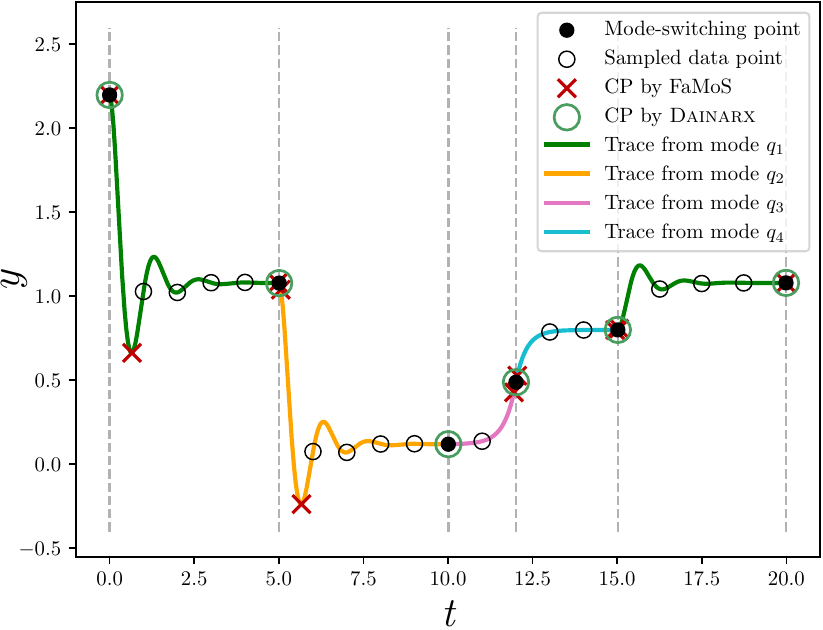}};
            \draw (-1.1, .5) node {\tiny \textcolor{customgreen}{$\traceDT_1$}};
            \draw (-0.22, -.36) node {\tiny \textcolor{customorange}{$\traceDT_2$}};
            \draw (.37, -.9) node {\tiny \textcolor{custompink}{$\traceDT_3$}};
            \draw (0.82, -.34) node {\tiny \textcolor{customcyan}{$\traceDT_4$}};
            \draw (1.52, -.1) node {\tiny \textcolor{customgreen}{$\traceDT_5$}};
        \end{tikzpicture}
        \caption{Comparison on \benchmark{loop\_syst}.}
        \label{fig:case_seg_cluster}
    \end{minipage}
    % \caption{Case studies demonstrating the effectiveness of {\ourmethod}.}
    % \label{fig:your_label}
\end{figure}

The upper part of \Cref{tab:merged_table} reports the %experimental 
results on inferring hybrid automata with linear dynamics and guards. In this category, {\ourmethod} exhibits a distinct advantage in effectiveness: It achieves the \emph{(tied-)highest accuracy} in detecting mode-switching (signified by $\hdt$) in all 19 benchmarks; It also attains the \emph{highest trace fidelity} (i.e., minimal discrepancies between inferred traces and ground-truth traces, signified by $\diffmax$ and $\diffavg$) across all benchmarks except for a comparable result with LearnHA for the \benchmark{cell} example. 
%In this case, \ourmethod has the same maximal discrepancy as LearnHA and is only 0.0001 lower in average discrepancy. 
% \Cref{fig:complex_tank,fig:buck_converter} 
\Cref{fig:complex_tank} depicts a more intuitive comparison of the performance on fitting testing traces. 
% In general, our experimental results demonstrate \ourmethod’s marked superiority over both LearnHA and FaMoS in accuracy and consistency. 

In terms of applicability, {\ourmethod} suffices to infer the underlying HA for \emph{all} the linear benchmarks with considerably accurate trace approximations. In contrast, LearnHA does not apply to high-order dynamics (with $k>1$), which is a crucial feature in modeling real-world dynamical systems. FaMoS fails in some benchmarks because of the presence of resets or inputs (\benchmark{tanks}, \benchmark{ball}, \benchmark{underdamped-c}) or implementation instability in clustering (the rest benchmarks marked by \enquote{--}). RIHAND fails in two complex benchmarks: \benchmark{dc\_motor} with 4th-order ODEs and \benchmark{underdamped-c} with 2nd-order ODEs, resets, and inputs.

% \vspace*{-2mm}
\subsubsection{Inference of Nonlinear Hybrid Systems}\label{subsec:experiments-nonlinear}
Next, we report the experimental results on learning hybrid automata with nonlinear dynamics and/or guards. In this category, we exclude FaMoS and
LearnHA due to the following reasons:
\begin{enumerate*}[label=(\roman*)]
\item FaMoS is dedicated to learning \emph{linear} hybrid systems; and
\item In principle, LearnHA supports polynomial ODEs and guard conditions. However, most of the LearnHA benchmarks (4 out of 5) in \cite{DBLP:conf/atva/GurungWS23} involve linear systems whilst the only nonlinear case lacks implementation details in its documentation. When applied to our nonlinear benchmarks, LearnHA generates hybrid automata with \emph{spurious variables} absent from the ground-truth specifications, making it challenging to evaluate the trace fidelity.
%For instance, in the \benchmark{lotkaVolterra} case, a hybrid automaton with two variables, LearnHA inferred ODEs introducing unexplained auxiliary variables without reporting the elimination process. 
Furthermore, LearnHA does not cope with systems with transcendental functions (e.g., trigonometric dynamics) or high-order nonlinearities. We 
\end{enumerate*}
\noindent
were thus not able to obtain valid inference results for nonlinear cases using LearnHA.

The lower part of \Cref{tab:merged_table} summarizes \ourmethod's performance on the inference of nonlinear hybrid systems. Akin to the linear case, \ourmethod achieves fairly high accuracy in both detecting mode switching and approximating the ground-truth traces. For the latter, it consistently outperforms RIHAND (in this noise-free setting). These results demonstrate the general applicability of {\ourmethod} to high-order nonlinear systems beyond polynomials, e.g., rational and transcendental dynamics or guards. Notably, \ourmethod retains high fidelity even in regimes where FaMoS and LearnHA fail, showing its robustness in handling diverse and complex nonlinear systems.

\subsubsection{Rationale behind Segmentation and Clustering}\label{subsec:experiments-seg-clustering}
We study a closed-loop control system (\benchmark{loop\_syst}; see \ifarxiv \cref{sec:appendix-case-study}\else\cite[Appx.~F]{DBLP:journals/corr/abs-2507-16426}\fi) to illustrate why the segmentation and clustering in {\ourmethod} -- both based on NARX model fitting -- yield better results than derivative-
%based segmentation 
and trace-similarity-based methods.
% clustering. 
For \emph{segmentation}, as shown in \cref{fig:case_seg_cluster}, the derivative-based approaches, like FaMoS, identify spurious mode-switching points exhibiting drastic variations in shape yet fail to detect genuine mode-switching points inducing smooth or subtle transitions.\footnote{The implementation of FaMoS provides a mechanism to rule out changepoints based on the result of clustering. However, we did not apply this mechanism in this experiment as it has the side-effect of eliminating true mode-switching points.}
%Additionally, these methods typically rely on manually set thresholds, making it difficult to ensure the rationality of the threshold values. 
In contrast, \ourmethod succeeds in detecting \emph{all} switching points by analyzing intrinsic dynamics of the trace \emph{without} manually specified thresholds. For a fair comparison of \emph{clustering}, we feed the true switching points to both FaMoS and \ourmethod and evaluate their %clustering 
performance. %under this condition. 
We observe that our NARX model-fitting-based strategy achieves the correct clustering:
% \begin{align*}
%     \{\textcolor{customgreen}{\traceDT_1},\textcolor{customgreen}{\traceDT_5}\} \uuplus \{\textcolor{customorange}{\traceDT_2}\} \uuplus \{\textcolor{custompink}{\traceDT_3}\} \uuplus \{\textcolor{customcyan}{\traceDT_4}\}~.\qquad\qquad\qquad\qquad\qquad\qquad\qquad\qquad\qquad\qquad\quad\ 
% \end{align*}%
\begin{align*}
    \{\textcolor{customgreen}{\traceDT_1},\textcolor{customgreen}{\traceDT_5}\} 
    \uuplus 
    \{\textcolor{customorange}{\traceDT_2}\} 
    \uuplus 
    \{\textcolor{custompink}{\traceDT_3}\} 
    \uuplus 
    \{\textcolor{customcyan}{\traceDT_4}\}~.
\end{align*}%
% \par}
% \noindent
But the clustering given by FaMoS is $\{\textcolor{customgreen}{\traceDT_1},\textcolor{customorange}{\traceDT_2}\} \uplus \{\textcolor{custompink}{\traceDT_3}\} \uplus \{\textcolor{customcyan}{\traceDT_4}\} \uplus \{\textcolor{customgreen}{\traceDT_5}\}$. This is because the trace-similarity-based strategy \emph{erroneously} concludes that segments of similar shape ($\traceDT_1$ and $\traceDT_2$) are generated by the same dynamics and segments of different shapes ($\traceDT_1$ and $\traceDT_5$) are not.

\subsection{Efficiency of {\ourmethod}}\label{subsec:efficiency}

We now report the efficiency of {\ourmethod} against the other baselines. The last 4 columns of \Cref{tab:merged_table} list the execution time of different inference tools across all benchmarks (detailed timings for individual phases are provided in \ifarxiv \cref{sec:appendix-executiontime}\else\cite[Appx.~G]{DBLP:journals/corr/abs-2507-16426}\fi).
%. We also provide detailed comparisons against FaMoS and LearnHA in \cref{tab:detailed_time} (\cref{sec:appendix-executiontime}) since these two have the same pipeline as {\ourmethod}. 
%We omit the explicit timings for the steps of mode characterization and reset learning as they are negligible compared to the other three steps. 
We note that the comparison may not be fair since the tools run on different platforms (FaMoS and RIHAND in MATLAB; \ourmethod and LearnHA in Python).

\begin{table}[t]
  \centering
  \caption{Results on inferring HA 
  %with linear (upper part) and nonlinear (lower part) dynamics/guards 
  under varying measurement and process noise with standard deviation {$\sigma_{\mathit{meas}}$} and {$\sigma_{\mathit{proc}}$}, respectively.}
  \label{tab:benchmark_robust}
  \vspace*{-2mm}
  \setlength{\tabcolsep}{2pt} 
  %\scriptsize 
  \resizebox{\linewidth}{!}{
  \begin{tabular}{lcccrrrrrrrrr}
   \toprule
    \multicolumn{6}{c}{Benchmark Details} & \multicolumn{2}{c}{$\epsilon_{\textit{noisy}}$} & \multicolumn{2}{c}{$\epsilon_{\textit{clean}}$} & \multicolumn{2}{c}{Total Time (seconds)}\\
    \cmidrule(lr){1-6} \cmidrule(lr){7-8} \cmidrule(lr){9-10} \cmidrule(lr){11-12}
    Name & $\abs{\modes}$ & $\abs{\varsO}$ & $k$ & $\sigma_{\mathit{meas}}$ & $\sigma_{\mathit{proc}}$ & RIHAND & \textbf{\ourmethod} & RIHAND & \textbf{\ourmethod} & RIHAND & \textbf{\ourmethod} \\
    \midrule
    \benchmark{Bouncing Ball}~\cite{kochdumper2025robust} & 1 & 2 & 1 & 0.010 & 0.10 & 0.0093 & \textbf{0.0089} & \textbf{0.0143} & 0.0270 & 12.51 & \textbf{1.04} \\
    \benchmark{oci}~\cite{kochdumper2025robust} & 2 & 2 & 1 & 0.010 & 0.10 & 0.4989 & \textbf{0.0151} & 0.7944 & \textbf{0.0322} & 32.86 & \textbf{0.44} \\
    \benchmark{Simple Heating System}~\cite{kochdumper2025robust} & 2 & 1 & 1 & 0.050 & 0.00 & 0.0726 & \textbf{0.0540} & 0.0656 & \textbf{0.0561} & 14.06 & \textbf{0.41} \\
    \benchmark{Variable Heating System}~\cite{kochdumper2025robust} & 3 & 2 & 1 & 0.050 & 0.00 & \textbf{0.0428} & 0.1043 &\textbf{0.0733}  & 0.1456 & 9.56 & \textbf{0.75} \\
    \benchmark{Buck Converter}~\cite{kochdumper2025robust} & 3 & 2 & 1 & 0.010 & 0.00 & \textbf{0.0299} & 0.0651 & \textbf{0.0412} & 0.0755 & 41.02 & \textbf{1.21} \\
    \benchmark{Complex Tank System}~\cite{kochdumper2025robust} & 8 & 3 & 1 & 0.010 & 0.00 & 0.4474 & \textbf{0.2075} & 0.1090 & \textbf{0.0572} & 244.87 & \textbf{3.31} \\
    \benchmark{simple\_linear} & 2 & 2 & 1 & 0.050 & 0.10& \textbf{0.0043}&  0.0140 &  \textbf{0.0150} & 0.0954  & 30.23 & \textbf{1.25}  \\
    \benchmark{jumper}~\cite{DBLP:conf/rss/Borquez0C0B24} & 2 & 4 & 1 & 0.020 & 0.10 & 0.6469 & \textbf{0.0536} & 0.4975 & \textbf{0.0575} & 20.51 & \textbf{1.58} \\
    \benchmark{two\_tank} & 2 & 2 & 1 & 0.020 & 0.05 & \textbf{0.0205} & 0.0229 & \textbf{0.0224} & 0.0244 & 29.43 & \textbf{6.92} \\
    \benchmark{Super Complex Tank System} & 12 & 3 & 1 & 0.002 & 0.1 & 0.1806 & \textbf{0.0707} & 0.0620 & \textbf{0.0308} & 170.12 & \textbf{7.93} \\
    \midrule
    \benchmark{lander}~\cite{DBLP:conf/fm/ZhaoYZGZC14} & 2 & 4 & 1 & 0.020 & 0.10 & \textbf{0.0088} & 0.0456 & \textbf{0.0038} & 0.0181 & 39.41 & \textbf{1.30} \\
    \benchmark{oscillator}~\cite{DBLP:conf/adhs/RamdaniN09} & 2 & 2 & 1 & 0.005 & 0.10 & 0.0765 & \textbf{0.0299} & 0.0335 & \textbf{0.0210} & 21.11 & \textbf{0.53} \\
    \benchmark{simple\_non\_linear} & 2 & 1 & 1 & 0.010 & 0.10 & 7.1605 & \textbf{0.0117} & 0.1545 & \textbf{0.0639} & 153.24 & \textbf{8.28} \\
    \benchmark{spacecraft} & 2 & 4 & 1 & 0.050 & 0.10 & 0.0936 & \textbf{0.0039} & 0.0461 & \textbf{0.0012} & 43.57 & \textbf{1.79} \\
    \bottomrule

  \end{tabular}
  }
\end{table}

Overall, we observe that both FaMoS and LearnHA suffer from a significant computational overhead for segment clustering, which dominates their total execution time. LearnHA, in particular, exhibits prolonged clustering phases -- typically 2--3$\times$ slower than the other steps -- resulting in suboptimal scalability to large systems. In contrast, \ourmethod exhibits a clear advantage in efficiency and scalability: It solves most of the benchmarks (21/27) within 5 seconds. A notable exception is the nonlinear benchmark \benchmark{sys\_bio} (with $\abs{\varsO} = 9$), which takes 27.4 seconds due to the combinatorial complexity of resolving high-dimensional mode interactions. Except for \benchmark{two\_tank}, \ourmethod outperforms the baselines by a wide margin: 1.4--17$\times$ faster than LearnHA, 1--10$\times$ faster than FaMoS, and 5--42$\times$ faster than RIHAND. This demonstrates the potential of applying {\ourmethod} to infer complex real-world system models.

% \cref{tab:time} presents the runtime\footnote{FaMoS is implemented in MATLAB, while \ourmethod and LearnHA are implemented in Python.} on learning each case. Both FaMoS and LearnHA experience substantial clustering times, which significantly contribute to their overall learning times. This is particularly evident in LearnHA, where the clustering process is relatively lengthy compared to other phases of the workflow, resulting in an extended total execution time. In contrast, our implementation demonstrates superior time efficiency across the majority of the benchmarks, with the learning time typically remaining quite short. The only exception is the case of sys\_bio, a nonlinear hybrid system with nine outputs, where the learning time exceeds 10s due to the complexity of the system. For all other benchmarks, our implementation consistently achieves faster performance than the implementations of FaMoS and LearnHA. This suggests that our implementation is able to handle various scenarios more efficiently.

\subsection{Robustness of {\ourmethod}}\label{subsec:robust}

% Next, we examine the robustness of {\ourmethod} to noise in comparison to the RIHAND approach \cite{kochdumper2025robust} dedicated to inferring hybrid automata from noisy data.

% To enhance the applicability of {\ourmethod} in real-world noisy environments, we further developed a noise-robust extension termed Robust~\ourmethod. A systematic comparison was conducted between Robust~\ourmethod and the recent noise-robust method~\cite{kochdumper2025robust} (hereinafter referred to as RIHAND) in terms of both effectiveness and computational efficiency. %Robust~\ourmethod introduces two additional hyperparameters, \emph{resampling\_interval} and \emph{truncation\_size}, which require careful joint tuning along with the original thresholds. The complete set of optimized hyperparameters is summarized in \cref{sec:hyperparameters}.

% \vspace*{-2mm}
% \subsubsection{{\color{red}Effectiveness of Robust \ourmethod}}\label{subsec:effectiveRD}

\begin{figure}[t]
    \begin{subfigure}[b]{0.23\textwidth}
    \centering
        \centering
        \includegraphics[width=\textwidth]{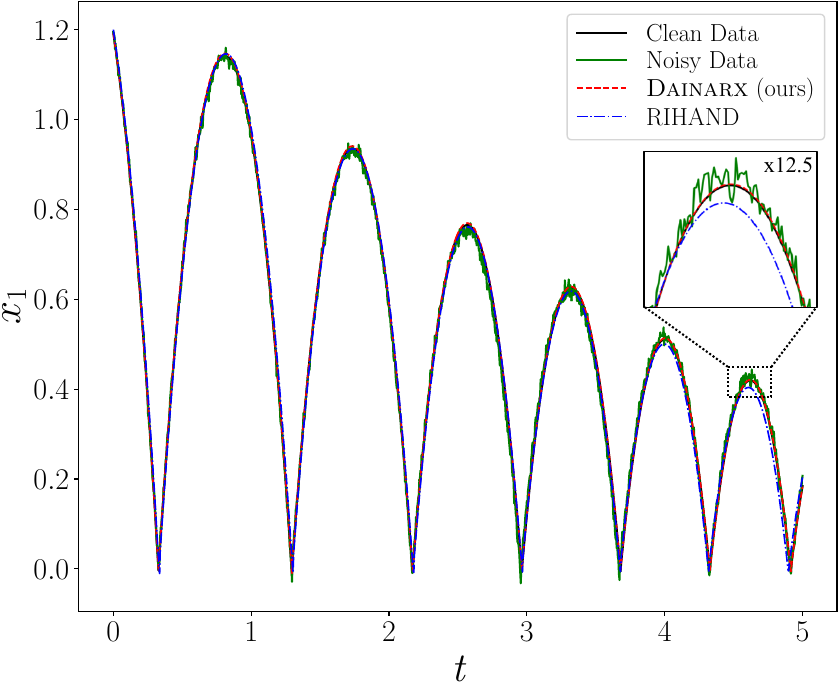}
        % \caption{trace fidelity for \benchmark{ball}}
        % \label{fig:noise_ball}
    \end{subfigure}
    \hfill
     \begin{subfigure}[b]{0.236\textwidth}
        \centering
        \includegraphics[width=\textwidth]{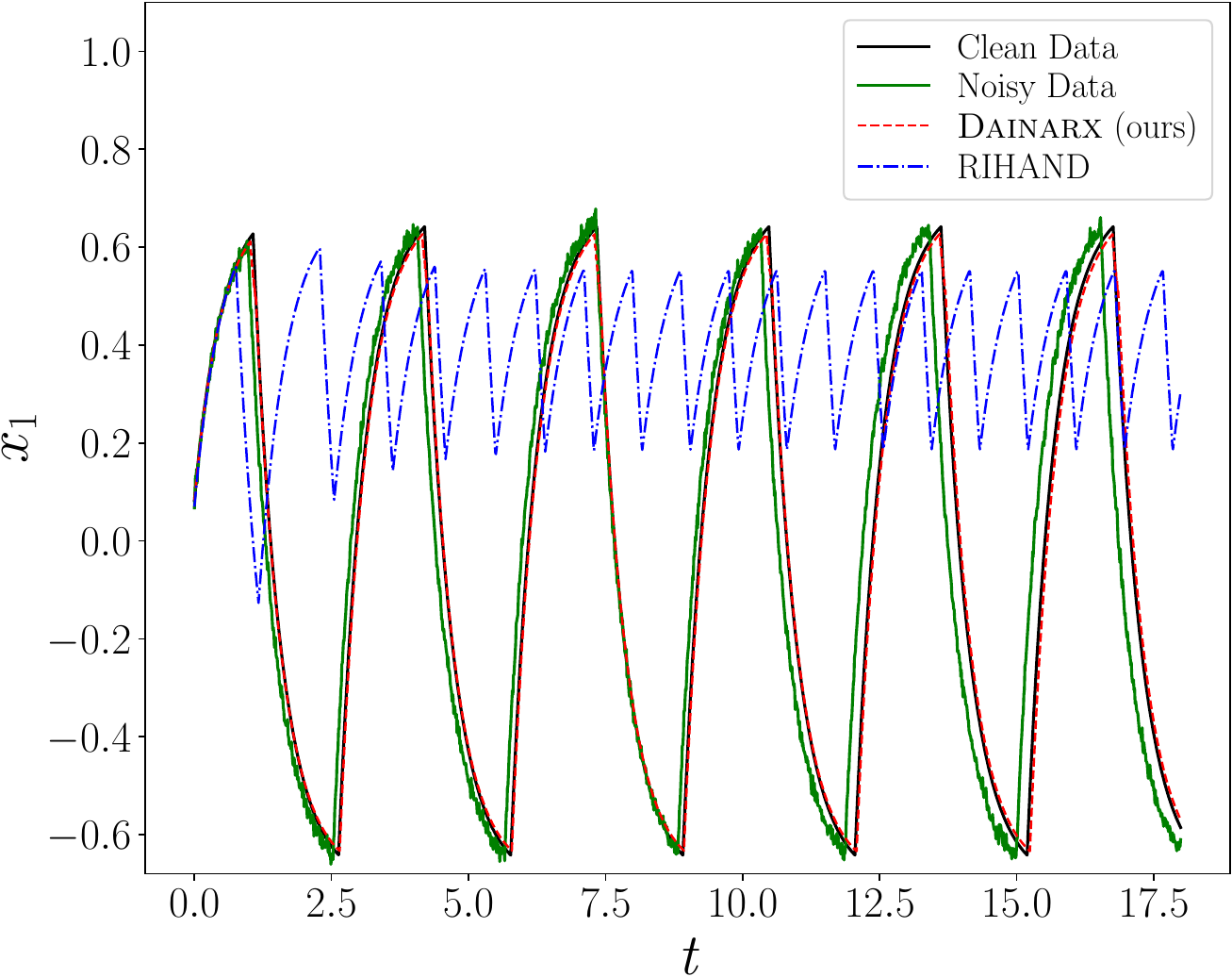}
        % \caption{trace fidelity for \benchmark{oci}}
        % \label{fig:noise_oci}
    \end{subfigure}
    \caption{Robustness of {\ourmethod} vs.\ RIHAND (left: \benchmark{ball}; right: \benchmark{oci}).}
    \label{fig:robustness-comparison}
\end{figure}

Next, we report the effectiveness and efficiency of \ourmethod in the noisy setting compared to the noise-oriented approach RIHAND \cite{kochdumper2025robust}.
%geared towards the inference of hybrid automata from noisy data.
In terms of trace fidelity, we adopt the following metrics:
\begin{align*}
    \epsilon_{\textit{noisy}} = \frac{\normDTWDisplay{\traceNARX(\tau),\traceDT_{\textit{noisy}}(\tau)}}{\sum^\traceDTLen_{i=0} \,\normDisplay{\traceDT_{\textit{noisy}}(\tau_i)}}~, 
    \ 
    \epsilon_{\textit{clean}} = \frac{1}{\traceDTLen} \!\cdot\! \sum^{\traceDTLen}_{i=0} \normOneDisplay{\!\traceNARX(\tau_i) - \traceDT_{\textit{clean}}(\tau_i)\!\!}
\end{align*}%
where ${\traceNARX}$ are predicted discrete-time traces, $\traceDT_{\textit{noisy}}$ are traces in $\data$ subject to both process and measurement noise, and $\traceDT_{\textit{clean}}$ are noise-free traces. Note that $\epsilon_{\textit{noisy}}$ is the metric used in RIHAND for measuring the discrepancy between the model-predicted trajectory and the \emph{noisy} data (leveraging the DTW distance \cite{berndt1994using}). Nonetheless, this metric does not reveal whether the learned model truly captures the underlying system dynamics or merely \emph{overfits the noisy data}. Therefore, we introduce the metric $\epsilon_{\textit{clean}}$ to quantify the averaged point-wise deviation of the predicted trajectory from the \emph{noise-free} trajectory.
\Cref{tab:benchmark_robust} shows that, overall, {\ourmethod} performs on par with RIHAND in terms of trace fidelity: {\ourmethod} achieves lower $\epsilon_{\textit{noisy}}$ on 9 out of 14 benchmarks and also lower $\epsilon_{\textit{clean}}$ on 8 of these 9 benchmarks; see \cref{fig:robustness-comparison} for a visualization on concrete benchmarks.

In terms of efficiency, {\ourmethod} is 1--2 orders of magnitude faster than RIHAND.
%, similar to the noise-free setting. 
A comparison between the noisy and noise-free results (\cref{tab:benchmark_robust} vs.\ \cref{tab:merged_table}) also shows that the resampling and TLS extensions of {\ourmethod} do not substantially degrade its efficiency. 
%These findings demonstrate the potential of scaling {\ourmethod} to real-world applications.

% A comparison between Table~\ref{tab:benchmark_robust} and Table~\ref{tab:detailed_time} in ~\cref{sec:appendix-executiontime} indicates that Robust~\ourmethod incurs higher computational cost compared to the original~\ourmethod due to the challenge of accurately detecting mode-switching points under noisy conditions, which affects both clustering and guard learning. Despite the additional resampling step, Robust~\ourmethod remains substantially faster than RIHAND with speedups ranging from approximately 4.25~$\times$ on \benchmark{two\_tank} to 74.68~$\times$ on \benchmark{oci}~(Table~\ref{tab:benchmark_robust}). These findings demonstrate that the proposed extension enhances noise robustness while preserving computational efficiency, supporting its applicability in real-world, resource-constrained settings.

% \input{sections/limitations}
% !TEX root = main.tex

\section{Related Work}\label{sec:related-work}

Hybrid automata inference has witnessed various techniques, including optimization \cite{DBLP:journals/ejcon/PaolettiJFV07}, data-driven feature extraction \cite{DBLP:conf/aaai/NiggemannSVMB12, DBLP:conf/icphys/LamraniBG18}, and machine learning \cite{DBLP:journals/eaai/HranisavljevicM20,vignolles2022hybrid}. 
%For instance, \cite{DBLP:journals/eaai/HranisavljevicM20} proposes a hierarchical deep learning framework using stacked restricted Boltzmann machines (RBMs) to abstract latent features from continuous traces, while \cite{vignolles2022hybrid} introduces a two-step approach combining DyClee clustering with support vector or random forest regression. 
% Despite progress, current approaches exhibit critical constraints. 
Despite progress, existing frameworks are restricted to linear or polynomial dynamics \cite{DBLP:journals/fac/JinAZZZ21,DBLP:conf/cdc/DayekhB024,DBLP:journals/tecs/SaberiFB23,DBLP:conf/hybrid/PlambeckBHF24, DBLP:journals/tcps/YangBKJ22,DBLP:conf/atva/GurungWS23,kochdumper2025robust}, switched systems \cite{DBLP:journals/fac/JinAZZZ21,DBLP:conf/cdc/DayekhB024,DBLP:journals/iandc/MadaryMAL22}, and switched (nonlinear) ARX models \cite{DBLP:journals/iandc/MadaryMAL22,DBLP:conf/hybrid/LauerB08,dayekh2024hybrid,DBLP:conf/hybrid/PlambeckBHF24}.
% precluding generalization to hybrid systems with guard conditions or resets
Learning general system models with, e.g., high-order dynamics, resets, inputs, and noise remains a challenge (recall \cref{tab:comparison}).
% is also a challenge to some of the previous work \cite{DBLP:journals/tecs/SaberiFB23}. 
% \cite{DBLP:conf/atva/GurungWS23} introduces type annotation to improve the inference of resets by utilizing domain knowledge. 
Our method suffices to infer hybrid automata with nonlinear ARX models, advancing the state-of-the-art (e.g., FaMoS \cite{DBLP:conf/hybrid/PlambeckBHF24}, LearnHA~\cite{DBLP:conf/atva/GurungWS23}, and HAutLearn \cite{DBLP:journals/tcps/YangBKJ22}) by admitting nonlinear hybrid automata with high-order, polynomial, rational, or trigonometric dynamics and/or guards under both process and measurement noise -- thus also amenable to robust HA identification \`{a} la \cite{kochdumper2025robust}. Experimental results demonstrated 
%the effectiveness, efficiency, and robustness of {\ourmethod}.
that {\ourmethod} significantly outperforms off-the-shelf inference tools for fitting clean data and performs on par with \cite{kochdumper2025robust} in the noisy setting. Moreover, {\ourmethod} provides partial guarantees on the correctness of mode-switching detection (\cref{subsec:segmentation}) and trace-segment clustering (\cref{subsec:clustering}). We note that stronger formal guarantees, such as soundness and completeness, have been established only for a highly restrictive class of models, e.g., constant ODEs by Soto et al.~\cite{garcia2019membership}.

\section{Conclusion, Limitation, and Future Work}

% \section{Conclusion}

% We have presented {\ourmethod} -- a derivative-agnostic framework for the inference of nonlinear hybrid automata. 

% This framework enables \emph{threshold-free} trace segmentation and clustering leveraging the technique of NARX model fitting throughout its workflow.
% % {\color{red}in noiseless situations and is able to identify through noisy traces}. 
{\ourmethod} is -- to the best of our knowledge -- the first approach that admits the inference of \emph{high-order non-polynomial dynamics} with \emph{exogenous inputs}, \emph{non-polynomial guard conditions}, and \emph{linear resets}. 
% Experimental results show that {\ourmethod} significantly outperforms state-of-the-art techniques for fitting clean data and performs on par with noise-oriented approaches in the noisy setting.
% %and {\color{red} \emph{noise}}. 
% %Future directions include extending {\ourmethod} to equip it with the PAC learning paradigm.

However, we observe the following limitations:
%of {\ourmethod} and provide potential solutions thereof.
%
First, {\ourmethod} requires -- in the nonlinear case -- user-provided nonlinear terms $\{f_i\}$ for the \emph{template} NARX model.
% to form the \emph{template} NARX model $\narx$. %Although 
Though we do not observe significant sensitivity of {\ourmethod}'s performance to the quality of templates, it still relies on a priori knowledge about the target system to obtain tight approximations. We foresee that machine-learning %models and 
techniques may be developed to \emph{learn a good NARX template} %$\narx$ 
from
% based on characteristics of 
the learning data.
%
% {\color{red}Second, {\ourmethod} can be generalized to accommodate noisy data; however, its effectiveness hinges on the existence of pronounced discrepancies in fitting errors between mode-switching and intra-mode points after resampling. Furthermore, since the sliding-window length scales proportionally with the resampling interval, the detection of closely spaced switching points becomes unreliable when their temporal separation falls below this expanded window. In addition, the fidelity of the learned trajectory gradually deteriorates under the sustained influence of noise. Hence, advancing robust hybrid-system identification, e.g., \cite{DBLP:conf/cdc/OzayLS09}, remains a critical direction. One promising avenue is to incorporate NARX-modeled dynamics with an explicit \emph{error term} $\epsilon[\tau]$, following \cite{DBLP:journals/iandc/MadaryMAL22,DBLP:conf/hybrid/LauerB08,DBLP:conf/cdc/LauerBV10}, and to extend {\ourmethod} accordingly.}
%
Second, {\ourmethod} admits a quantitative means to measure the discrepancy between the inferred model and the \emph{sampled learning data instead of the real system trajectories}. This is a common challenge for hybrid system identification as authentic system trajectories are often intractable to obtain, especially for nonlinear dynamics \cite{Gan18}. However, it would be interesting to explore whether techniques like \emph{probably approximately correct} (PAC) learning \cite{DBLP:conf/stoc/Valiant84} can be leveraged to establish quantitative guarantees.
%for the inferred hybrid automata, as is done for the learning of timed automata \cite{DBLP:conf/icfem/ShenAZZ0Z20}, symbolic automata \cite{DBLP:conf/birthday/MalerM17}, and dynamical systems~\cite{DBLP:conf/icml/AdigaKMRV19,DBLP:conf/icml/QiuAMRRSV24}.

%%
\begin{acks}
%% acks environment is optional
%% contents suppressed with 'anonymous'
   %% Commands \grantsponsor{<sponsorID>}{<name>}{<url>} and
   %% \grantnum[<url>]{<sponsorID>}{<number>} should be used to
   %% acknowledge financial support and will be used by metadata
   %% extraction tools.
%   This material is based upon work supported by the
%   \grantsponsor{GS100000001}{National Science
%     Foundation}{http://dx.doi.org/10.13039/100000001} under Grant
%   No.~\grantnum{GS100000001}{nnnnnnn} and Grant
%   No.~\grantnum{GS100000001}{mmmmmmm}.  Any opinions, findings, and
%   conclusions or recommendations expressed in this material are those
%   of the author and do not necessarily reflect the views of the
%   National Science Foundation.
%
This work is funded by the NSFC (No.\ 62572427), by the ZJNSF Major Program (No.\ LD24F020013), by the CASC Open Fund (No.\ LHCESET202502), and by the Fundamental Research Funds for the Central Universities of China (No.\ 226-2024-00140).
%and by the Zhejiang Key Laboratory Project (No.\ 2024E10001). 
%The authors would like to thank the anonymous reviewers for their constructive feedback.
\end{acks}

\bibliographystyle{ACM-Reference-Format}
\bibliography{references}

% \newpage
\ifarxiv
\appendix
% !TEX root = main.tex

\section*{Appendix}

\begin{table*}[!t]
  \centering
  \caption{Experimental results on the execution time (in seconds) of the inference tools.}
  \label{tab:detailed_time}
  \setlength{\tabcolsep}{2pt} 
  %\scriptsize 
  \resizebox{1\textwidth}{!}{
  \begin{tabular}{lcccrrrrrrrrrrrrr}
   \toprule
    \multicolumn{4}{c}{Benchmark Details}  & \multicolumn{3}{c}{Segmentation Time} & \multicolumn{3}{c}{Clustering Time} & \multicolumn{3}{c}{Guard-Learning Time} & \multicolumn{4}{c}{Total Time} \\
    \cmidrule(lr){1-4} \cmidrule(lr){5-7} \cmidrule(lr){8-10} \cmidrule(lr){11-13}\cmidrule(lr){14-17}
    Name & $\abs{\modes}$ & $\abs{\varsO}$ & $k$ & LearnHA & FaMoS & \textbf{\ourmethod} & LearnHA & FaMoS & \textbf{\ourmethod} & LearnHA & FaMoS & \textbf{\ourmethod} & LearnHA & FaMoS & RIHAND& \textbf{\ourmethod} \\
    \midrule
    \benchmark{buck\_converter}~\cite{DBLP:conf/hybrid/PlambeckBHF24} & 3 & 2 & 1 & 2.10 & \textbf{0.12} & 1.06 & 25.60 & 3.28 & \textbf{0.93} & 0.95 & 0.03 & \textbf{0.02} & 28.70 & 4.43 & 31.07 &\textbf{2.12} \\
    \benchmark{complex\_tank}~\cite{DBLP:conf/hybrid/PlambeckBHF24} & 8 & 3 & 1 & 4.65 & \textbf{0.53} & 1.70 & 34.00 & 18.40 & \textbf{3.46} & 1.84 & \textbf{0.03} & 0.34 & 40.50 &  19.00 & 161.03&\textbf{5.73} \\
    \benchmark{multi\_room\_heating}~\cite{DBLP:conf/hybrid/PlambeckBHF24} & 4 & 3 & 1 & 2.90 & \textbf{0.35} & 1.12 & 19.12 & 7.71 & \textbf{1.57} & 1.22 & \textbf{0.03} & 0.12 &  23.30 & 8.12 & 63.29 & \textbf{2.96} \\
    \benchmark{simple\_heating\_syst}~\cite{DBLP:conf/hybrid/PlambeckBHF24} & 2 & 1 & 1 & 2.09 & \textbf{0.26} & 0.52 & 10.70 & 8.10 & \textbf{0.19} & 0.58 & \textbf{0.01} & \textbf{0.01} & 13.40 & 8.39 & 15.23 & \textbf{0.78} \\
    \benchmark{three\_state\_HA}~\cite{DBLP:conf/hybrid/PlambeckBHF24} & 3 & 1 & 2 & -- & \textbf{0.14} & 0.55 & -- & 1.30 & \textbf{0.31} & -- & 0.01 & \textbf{0.00} & -- & 1.45 &37.40 &  \textbf{0.92} \\
    \benchmark{two\_state\_HA}~\cite{DBLP:conf/hybrid/PlambeckBHF24} & 2 & 1 & 2 & -- & \textbf{0.20} & 0.54 & -- & 1.50 & \textbf{0.22} & -- & \textbf{0.01} & \textbf{0.01} & -- & 1.72 &25.14 & \textbf{0.80} \\
    
    \benchmark{variable\_heating\_syst}~\cite{DBLP:conf/hybrid/PlambeckBHF24} & 3 & 2 & 1 & 2.10 & \textbf{0.29} & 0.95 & 8.55 & 3.93 & \textbf{0.65} & 0.65 & \textbf{0.01} & 0.11 & 11.30 & 4.23 &31.75 & \textbf{1.79} \\
    
    \benchmark{cell}~\cite{DBLP:conf/atva/GurungWS23} & 4 & 1 & 1 & 9.28 & \textbf{0.44} & 3.38 & 55.10 & 39.94 & \textbf{3.80} & 0.80 & \textbf{0.02} & \textbf{0.02} & 65.20 & 40.42 &78.47& \textbf{7.47} \\
    
    \benchmark{oci}~\cite{DBLP:conf/atva/GurungWS23} & 2 & 2 & 1 & 2.14 & -- & \textbf{0.93} & 5.83 & -- & \textbf{0.36} & 0.48 & -- & \textbf{0.03} & 8.46 & -- & 23.81 & \textbf{1.41} \\
    
    \benchmark{tanks}~\cite{DBLP:conf/atva/GurungWS23} & 4 & 2 & 1 & 2.21 & -- & \textbf{1.17} & 16.98 & -- & \textbf{1.50} & 1.56& -- & \textbf{1.04} & 20.79 & -- & 19.33 & \textbf{3.82} \\
    
    \benchmark{ball}~\cite{DBLP:conf/atva/GurungWS23} & 1 & 2 & 1 & 0.59 & -- & \textbf{0.26} & 2.17 & -- & \textbf{0.13} & 0.62 & -- & \textbf{0.31} & 3.38 & -- & 9.95 &\textbf{0.72} \\
    
    \benchmark{dc\_motor} & 2 & 2 & 4 & -- & -- & \textbf{1.50} & -- & -- & \textbf{2.09} & -- & -- & \textbf{0.05} & -- & -- & --& \textbf{4.08} \\
    
    \benchmark{simple\_linear} & 2 & 2 & 1 & 2.23 & \textbf{0.19} & 0.60 & 9.99 & 3.58 & \textbf{0.86} & 0.56 & \textbf{0.02} & 0.42 & 12.79 & 3.80 & 25.87 & \textbf{1.97} \\
    
    \benchmark{jumper}~\cite{DBLP:conf/rss/Borquez0C0B24} & 2 & 4 & 1 & 0.62 & -- & \textbf{0.11} & 2.24 & -- & \textbf{0.11} & \textbf{0.49} & -- & 1.78 & 3.35 & -- &13.90 & \textbf{2.38} \\

    \benchmark{loop\_syst} & 4 & 2 & 2 & -- & \textbf{0.20} & 1.36 & -- & 3.09 & \textbf{1.57} & -- & \textbf{0.01} & \textbf{0.01} & -- &  3.30 & 77.34 & \textbf{3.10} \\
    
    \benchmark{two\_tank} & 2 & 2 & 1 & 2.04 & 0.24 & \textbf{0.18} & 4.97 & 2.69 & \textbf{0.18} & 0.41 & \textbf{0.01} & 3.20 & 7.42 &  \textbf{2.94} & 24.16 &4.61 \\
    
    \benchmark{underdamped} & 2 & 2 & 2 & -- & -- & \textbf{1.42} & -- & -- & \textbf{0.67} & -- & -- & \textbf{0.03} & -- & -- & 77.98& \textbf{2.25} \\
    \benchmark{underdamped-c} & 4 & 2 & 2 & -- & -- & \textbf{1.52} & -- & -- & \textbf{2.65} &  --& -- & \textbf{0.06}  & -- &  --& -- &\textbf{4.52}  \\ 
    \benchmark{super\_complex\_tank} & 12 & 3 & 1 & 4.38 & \textbf{0.87} & 1.75 & 34.56 & 17.38 & \textbf{3.21} & 2.41 & 0.38 & \textbf{0.29} & 41.39 & 18.66 & 127.68 & \textbf{5.43} \\
    \midrule
    \benchmark{lander}~\cite{DBLP:conf/fm/ZhaoYZGZC14} &2 &4&1 & \multirow{8}{*}{--} & \multirow{8}{*}{--} & \textbf{1.38} &  \multirow{8}{*}{--}& \multirow{8}{*}{--} & \textbf{1.03} & \multirow{8}{*}{--} & \multirow{8}{*}{--} &  \textbf{0.00} & \multirow{8}{*}{--} & \multirow{8}{*}{--} & 26.53& \textbf{2.58} \\
    \benchmark{lotkaVolterra}~\cite{DBLP:conf/arch/GerettiSABCFILM23} &2&2& 1&& & \textbf{0.44} & & &\textbf{0.24} & && \textbf{0.02} & &&  32.49&\textbf{0.76} \\
    \benchmark{simple\_non\_linear} &2&1&1& && \textbf{3.04} & && \textbf{2.28} & & &\textbf{0.02} & && 53.71&\textbf{5.70} \\
    \benchmark{simple\_non\_poly} &2&1&1& && \textbf{2.46} & && \textbf{1.20} & && \textbf{0.03} & &&78.78& \textbf{3.96} \\
    \benchmark{oscillator}~\cite{DBLP:conf/adhs/RamdaniN09} &2&2&1& && \textbf{0.65} & && \textbf{0.43} && & \textbf{0.01} && & 28.87&\textbf{1.13} \\
    \benchmark{spacecraft}~\cite{DBLP:conf/cpsweek/ChanM17} &2&4&1&& & \textbf{1.20} & && \textbf{0.72} & && \textbf{0.01} & &&38.96& \textbf{2.01} \\
    \benchmark{sys\_bio}~\cite{DBLP:journals/iandc/WangCXZK22} &2&9&1& && \textbf{13.17} & & & \textbf{12.60} & && \textbf{0.20} & &&239.86& \textbf{27.40} \\
    \benchmark{duffing} &2&1&2& && \textbf{3.92} & && \textbf{2.98} & && \textbf{0.06} & && 150.93&\textbf{7.34} \\
    \bottomrule
  \end{tabular}
  }
\end{table*}

\section{Proofs of Theorems}\label{sec:appendix-proofs}

\restateExactFitting*

\begin{proof}
    We first show that $E_S=0$ and $N \models S$ are equivalent. Since $E_i = \norm{\cdot} \geq 0$, $E_S = 0$ implies $E_i=0$ for any $i$ and $\Omatrix_{i,:} - \cmatIns_{i,:} \cdot \Dmatrix_i$ is the corresponding form for variable $x_i$ in \cref{eq:narx-templated}. Namely, $E_S=0$ indicates that \cref{eq:narx-templated} is true for every variable, i.e. $N \models S$. Next, we show that $E_S=0$ if and only if $S$ is fittable by $\narx$. Since the above proof shows that $E_S=0$ yields an instance $\narxIns \models S$, we only need to prove that $E_S=0$ when $S$ is fittable by $\narx$. For any instance $N' = \narx[\cmatIns'] \in \narxInsSet$ such that $N'\models S$, by substituting $\cmatIns'$ for $\cmatIns$ in \cref{eqn:optimization}, we get $E_S=0$ according to \cref{eq:narx-templated}.
    This completes the proof.
\end{proof}

\begin{lemma}[Monotonicity of Fittability]
    \label{lma:fittable-monotony}
    Given a template NARX model $\narx$ and two traces $\traceDT, \traceDT'$, where $\traceDT'$ is a segment of $\traceDT$. If $\traceDT$ is fittable by $\narx$, then $\traceDT'$ is fittable by $\narx$. 
\end{lemma}

\begin{proof}
    Since $\traceDT$ is fittable by $\narx$, according to \cref{def:trace-fitting}, there exists $N \in \narxInsSet$ that \cref{eq:narx-templated} holds for all $\tau$. Since $\traceDT'$ is a segment of $\traceDT$, \cref{eq:narx-templated} holds for all $\tau$ also for $\traceDT'$, i.e., $\traceDT'$ is fittable by $\narx$.
\end{proof}

\begin{lemma}[Uniqueness of CP]
    \label{lma:cp-unique}
    Given a template NARX model $\narx$ and a trace $\traceDT$. The set of changpoints defined by \cref{property:changepoints} is unique.
\end{lemma}
\begin{proof}
    
    We have $CP = \{p_0, p_i,... p_s\}$ where $p_i$ satisfies the well-ordered relation: $p_0 = 0 < p_1 < \ldots p_i < \ldots  < p_s = \traceDTLen + 1$. We prove the claim by induction on $i$.
    For the base case, we have $p_0 = 0$ is unique; For the induction step, assume $p_i$ is unique, we prove by contradiction: Suppose $p_{i+1}$ is not unique, i.e., there exists $p_{i+1} < p'_{i+1}$ satisfying that both $\traceDT_{p_i, p_{i+1}}$ and $\traceDT_{p_i, p'_{i+1}}$ are fittable by $\narx$, but $\traceDT_{p_i, (p_{i+1})+1}$ is not fittable by $\narx$. Since $\traceDT_{p_i, (p_{i+1})+1}$ is a segment of $\traceDT_{p_i, p'_{i+1}}$, according to \cref{lma:fittable-monotony}, $\traceDT_{p_i, (p_{i+1})+1}$ is fittable by $\narx$ leading to a contradiction. This completes the proof.
\end{proof}

\begin{lemma}[CP is Subset]
    \label{lma:cp-unique}
    Given a template NARX model $\narx$ and a trace $\traceDT$. The set of changpoints defined by \cref{property:changepoints} is unique.
\end{lemma}
\begin{proof}
    
    We have $CP = \{p_0, p_i,... p_s\}$ where $p_i$ satisfies the well-ordered relation: $p_0 = 0 < p_1 < \ldots p_i < \ldots  < p_s = \traceDTLen + 1$. We prove the claim by induction on $i$.
    For the base case, we have $p_0 = 0$ is unique; For the induction step, assume $p_i$ is unique, we prove by contradiction: Suppose $p_{i+1}$ is not unique, i.e., there exists $p_{i+1} < p'_{i+1}$ satisfying that both $\traceDT_{p_i, p_{i+1}}$ and $\traceDT_{p_i, p'_{i+1}}$ are fittable by $\narx$, but $\traceDT_{p_i, (p_{i+1})+1}$ is not fittable by $\narx$. Since $\traceDT_{p_i, (p_{i+1})+1}$ is a segment of $\traceDT_{p_i, p'_{i+1}}$, according to \cref{lma:fittable-monotony}, $\traceDT_{p_i, (p_{i+1})+1}$ is fittable by $\narx$ leading to a contradiction. This completes the proof.
\end{proof}

\restateSegmentation*

\begin{proof}
    In the absence of the three corner cases, we have that the changepoint interval is greater than $w$, and the segment between two changepoints are fittable by $\narx$. We also know that $\{\traceDT,\traceDT'\}$is fittable by $\narx$ when (i) the overlap of two segments $\traceDT, \traceDT'$ is greater than $w-1$, and (ii) $\traceDT, \traceDT'$ are fittable by $\narx$. Let the changepoints obtained by \cref{alg:segmentation-sliding} be $CP' = \{p'_0, p'_1,... ,p '_s\}$ and the correct changepoint be $CP=\{p_0,p_1,... p_c\}$ (which is unique according to \cref{lma:cp-unique}). We now prove that $CP' = CP$ via mathematical induction: 
    For the base case, we have $p'_0 = p_0 = 0$; For the induction step, assume $p'_i = p_i$, we first prove that $\traceDT_{p_i.p'_{i+1}}$ is fittable by $\narx$. Since the interval of changpoints is greater than $w$, $\traceDT_{p_i,p_i+w}$ is fittable by $\narx$. Now we need to prove that for every $h=0, 1, \ldots p'_{i+1} - w$, $\traceDT_{p_i, p_i+h+w}$ is fittable by $\narx$. We again use induction on $h$. First, it is proved for $h=0$; Second, when $\traceDT_{p_i, p_i+h+w}$ is fittable by $\narx$, according to \cref{alg:segmentation-sliding}, we have $\traceDT_{p_i + h + 1, p_i+h+w+1}$ is fittable by $\narx$, thus $\{\traceDT_{p_i, p_i+h+w}, \traceDT_{p_i + h + 1, p_i+h+w+1}\}$ is fittable by $\narx$, which implies the fittability of $\traceDT_{p_i,p_i+h+w+1}$ (as $w > k + 1$) by definition. 
    % as and all \cref{eq:narx-templated} of $\traceDT_{p_i,p_i+h+w+1}$ are contained by it (because $w > k + 1$). So $\traceDT_{p_i,p_i+h+w+1}$ is fittable by $\narx$. 
    Hence, by induction, $\traceDT_{p_i.p'_{i+1}}$ is fittable by $\narx$. Now, suppose $p'_{i+1} \neq p_{i+1}$, we do a case distinction:
    \begin{enumerate*}[label=(\roman*)]
        \item For $p'_{i+1} > p_{i+1}$, we have that $\traceDT_{p_i,(p_{i+1})+1}$ is not fittable by $\narx$ and $\traceDT_{p_i,(p_{i+1})+1}$ is a segment of $\traceDT_{p_i,p'_{i + 1}}$. According to \cref{lma:fittable-monotony}, $\traceDT_{p_i,(p_{i+1})+1}$ is fittable by $\narx$, which leads to a contradiction;
        \item For $p'_{i+1} < p_{i+1}$, according to \cref{alg:segmentation-sliding}, $\traceDT_{(p'_{i+1}) - w + 1, (p'_{i+1})+1}$ is not fittable by $\narx$, but $\traceDT_{(p'_{i+1}) - w + 1, (p'_{i+1})+1}$ is a segment of $\traceDT_{p_i, p_{i+1}}$, a contradiction follows by applying \cref{lma:fittable-monotony}.
    \end{enumerate*}
    Hence, $p'_{i+1}$ is equal to $p_{i+1}$ and we have that $CP'=CP$.
\end{proof}

% \begin{comment}
\section{Trace Segmentation via Binary Search}\label{sec:appendix-algorithms}

\begin{algorithm}[h]
    \caption{Trace Segmentation Based on NARX Model Fitting (via binary search)}
    \label{alg:segmentation-binary}
    \SetKwInput{Input}{Input}\SetKwInOut{Output}{Output}\SetNoFillComment
    \Input{A discrete-time trace $\traceDT$ of length $\traceDTLen+1$ %per \cref{eq:learning-data} 
    and a template NARX model $\narx$ of order $k \in \Nats$.}
    \Output{A set $\segments_{\traceDT}$ of trace segments of $\traceDT$ under $\narx$.}
    $\cpoints \gets \{0\}$;\ \ $\lastcp \gets 0$ \tcp*{$\mathtt{initialization}$}
    % \DontPrintSemicolon
    \While(\tcp*[f]{$\mathtt{find\;changepoints\;via\;binary\;search}$}){$\lastcp < \traceDTLen + 1$}{
        $\lb \gets \lastcp + k + 1$;\ \ $\rb \gets \traceDTLen + 1$ \tcp*{$\mathtt{initialize\;bounds}$}
        % \tcp*{initialize lower- and upper-bounds}
        \While{$\lb < \rb$} {
            $\midElement \gets \left\lceil \frac{\lb + \rb}{2} \right\rceil$\;
            \If(\tcp*[f]{$\mathtt{potential\;changepoint}$}){$\traceDT_{\lastcp,\midElement}$ is not fittable by $\narx$}{
                $\rb \gets \midElement - 1$\;
            }
            \Else{
                $\lb \gets \midElement$\;
            }
        }
        $\cpoints \gets \cpoints \cup \{\rb\}$;\ \ $\lastcp \gets \rb$ \tcp*{$\mathtt{update\;the\;changepoints}$}
    }
    \tcc{$\mathtt{segment}\;\traceDT\;\mathtt{as\;per}\;\cpoints$}
    $\segments_{\traceDT} \gets \{\traceDT_{p_0, p_1}, \traceDT_{p_1, p_2}, \ldots, \traceDT_{p_{s-1}, p_s}\} \quad \text{for} \quad \cpoints = \{p_0, p_1,\ldots,p_s\}$ %\tcp*{segment $\traceDT$ as per $\cpoints$}
    
    \smallskip
    \For{$\segment \in \segments$\label{alg:segmentation-binary-check}}{
        \If{$\segment$ is not fittable by $\narx$\label{alg:segmentation-binary-nonfittable}}{
            $\segments_{\traceDT} \gets \segments_{\traceDT} \setminus \{\segment\}$ \tcp*{$\mathtt{drop\;non\!-\!fittable\;segments\;in}\;\segments_{\traceDT}$}  \label{alg:segmentation-binary-drop}
        }
    }
    \Return $\segments_{\traceDT}$\;
\end{algorithm}

% \bigskip
This algorithm exhibits a (logarithmic) increase in time complexity and, as observed in practice, the binary search-based segmentation relies on high-quality template NARX models. Therefore, we opt for the sliding window-based implementation, which yields considerably accurate models even in the absence of high-quality templates, as demonstrated in \cref{subsec:efficiency}.

Note that, the \emph{posterior screening} in Lines \ref{alg:segmentation-binary-check} to \ref{alg:segmentation-binary-drop} of \cref{alg:segmentation-binary} ensures that every trace segment in the eventually obtained set $\segments_{\traceDT}$ is fittable by the template NARX model $\narx$. Such a screening process is necessary due to the following two \emph{corner cases}:
\begin{enumerate*}[label=(\roman*)]
    \item the provided template NARX model $\narx$ does not suffice to fit some of the trace segments; and
    \item the distance between some neighboring mode-switching points falls below the order $k$ of $\narx$.
\end{enumerate*}
In the absence of these two corner cases, \cref{alg:segmentation-binary} is correct:
% we can establish the correctness of \cref{alg:segmentation-binary} as follows:

\begin{theorem}[Correctness of Segmentation in \cref{alg:segmentation-binary}]\label{thm:segmentation-binary}
Given a discrete-time trace $\traceDT$ and a template NARX model $\narx$. The set $\segments_{\traceDT}$ of trace segments returned by \cref{alg:segmentation-binary} -- in the absence of the above two corner cases -- satisfies the changepoint property as stated in \cref{property:changepoints}.
\end{theorem}

\begin{proof}
    The proof is similar to that of \cref{thm:segmentation-sliding}. We need to show that the set of changepoints $ CP'$ obtained by \cref{alg:segmentation-binary} coincides with $CP$. We prove again by induction. For the base case, we have $p'_0 = p_0= 0$; For the induction step, assume $p'_i = p_i$, we prove that $p_{i+1}$ lies at the bisection boundary, which means that $p'_{i+1}$ = $p_{i+_1}$. For any $p_g > p_i+1$, since $\traceDT_{p_i, (p_{i+1})+1}$ is a segment of $\traceDT_{p_i,p_g}$, if $\traceDT_{p_i,p_g}$ is fittable by $\narx$, by \cref{lma:fittable-monotony}, $\traceDT_{p_i, (p_{i+1})+1}$ is fittable by $\narx$, which contradicts \cref{property:changepoints}. Thus, $\traceDT_{p_i,p_g}$ is not fittable by $\narx$. For any $p_i+ k < p_l < p_{i+1}$, since $\traceDT_{p_i,p_{i+1}}$ is fittable by $\narx$ and $\traceDT_{p_i,p_l}$ is a segment of $\traceDT_{p_i,p_{i+1}}$, it follows from \cref{lma:fittable-monotony} that $\traceDT_{p_i,p_l}$ is fittable by $\narx$. Therefore, fittability satisfies the dichotomy property and $p_{i+1}$ lies at the dichotomy boundary. Thus, $p'_{i+1}$ obtained by the binary search-based algorithm is equal to $p_{i+1}$.
    This completes the proof.
\end{proof}
% \end{comment}

\section{The Minimally Mergeable Criterion}\label{sec:appendix-mmergeable}

The notion of mergeable in \cref{def:mergeable} is simple; it guarantees that all %trace 
segments generated by the same %system 
mode are mergeable. However, the reverse does not hold: The set of mergeable segments may actually be generated by \emph{different} mode dynamics:
%demonstrated by the following example.

\begin{example}[Aggressive Merging]\label{example:aggressive-merging}
Consider four trace segments with a single output variable $x$:
% \begin{align*}
%     \traceDT_1 &\eeq \langle (0,1), (1,2), (2,3), (3,4), (4,5)\rangle~,  &\traceDT_3 &\eeq \langle (0,1), (1,\mathrm{e}), (2,\mathrm{e}^2), (3,\mathrm{e}^3), (4,\mathrm{e}^4)\rangle,\\
%     \traceDT_2 &\eeq \langle (0,1), (1,3), (2,5), (3,7), (4,9)\rangle~,  &\traceDT_4 &\eeq \langle (0,0), (1,0), (2,0), (3,0), (4,0)\rangle~.
% \end{align*}%
\begin{align*}
    \traceDT_1 &= \langle (0,1), (1,2), (2,3), (3,4)\rangle~,  &\traceDT_3 &= \langle (0,1), (1,\mathrm{e}), (2,\mathrm{e}^2), (3,\mathrm{e}^3)\rangle,\\
    \traceDT_2 &= \langle (0,1), (1,3), (2,5), (3,7)\rangle~,  &\traceDT_4 &= \langle (0,0), (1,0), (2,0), (3,0)\rangle~.
\end{align*}%
Observe that $\traceDT_1$ and $\traceDT_2$ may be generated (i) by the same second-order mode dynamics $x^{(2)} = 0$; or (ii) by different first-order mode dynamics of $x^{(1)} = 1$ and $x^{(1)} = 2$, respectively. However, the mergeable criterion \`{a} la \cref{def:mergeable} applies only to the former case: $\traceDT_1$ and $\traceDT_2$ are mergeable because they are fittable by the same second-order NARX model \(x[\tau] = 2x[\tau - 1] - x[\tau - 2]\). The same argument holds for trace segments $\traceDT_3$ and $\traceDT_4$: they may be generated (i) by the same first-order mode dynamics $x^{(1)} = x$; or (ii) by different mode dynamics of $x^{(1)} = x$ (of order 1) and $x^{(0)} = 0$ (of order 0), respectively. However, the mergeable criterion applies only to the former case: $\traceDT_3$ and $\traceDT_4$ are mergeable because they are fittable by the same first-order NARX model \(x[\tau] = \mathrm{e} x[\tau - 1]\).
\qedT
\end{example}
% Both of these signals can be fitted by the second-order difference equation \(x[\tau] = 2x[\tau - 1] - x[\tau - 2]\), so they are mergeable. However, they are not minimally mergeable because they can each be fitted by a first-order difference equation, but not both at the same time. For the original hybrid system, they may come from the same mode \(x'' = 0\) or from two separate modes \(x' = 1\) and \(x' = 2\).

% \begin{example}
%     Given a difference template \(\tilde{\narx} = (1, \{x\}, \emptyset)\), there are two traces:
%     \begin{equation}
%         \begin{split}
%             t_1 &= \{ [1, e, e^2, e^3, e^4, \dots] \} \\
%             t_2 &= \{ [0, 0, 0, 0, 0, \dots] \}
%         \end{split}
%         \nonumber
%     \end{equation}
    
%     Both of these signals can be fitted by the first order difference equation \(x[\tau] = ex[\tau - 1]\) and they are also mergeable. But since the signals in \(t_2\) can also be fitted by the 0th order difference equation, they are not minimally mergeable either. Similarly, in the original hybrid system, they may both come from mode \(x' = x\), or they may come from modes \(x' = x\) and \(x = 0\), respectively.
% \end{example}

{\ourmethod} provides a less aggressive merging criterion called \emph{minimally mergeable}:
\begin{definition}[Minimally Mergeable Traces]\label{def:min-mergeable}
%Given a discrete-time trace $\traceDT$ and a template NARX model $\narx$. 
Given a set $S$ of discrete-time trace segments, let $\hat{k}_S \defeq \min\{k \in \Nats\mid k~\text{is %the
order of}~\narxIns \in \narxInsSet~\text{such that}~\narxIns \models S\}$ be the minimal order of NARX models that fit $S$ under template $\narx$. Then, $S$ is \emph{minimally mergeable} w.r.t.\ $\narx$ if $\hat{k}_{S} = \hat{k}_{\{\traceDT\}}$ for any $\traceDT \in S$.
% Given a discrete-time trace $\traceDT$ and a template NARX model $\narx$. Let $\segments_{\traceDT}$ be the set of segments obtained  of trace segments returned by \cref{alg:segmentation-sliding} 
%     \(\tilde{\narx} = (k, X, f)\) is a difference template, \(t_1, t_2\) are two traces (sets of signals), and there exists two system of difference equations \(\narx_1,\narx_2\) based on \(\tilde{\narx}\) satisfying \(\narx_1 \models t_1, \narx_2 \models t_2\), then:
%     \begin{itemize}
%         \item Call \(t_1, t_2\) mergeable when \(\exists \narx = (k, X, f, \mathcal{A}),\narx \models t_1, t_2\)
%         \item Let \(k_1'\) be the smallest \(k_1\) of all satisfactions of \(\exists \narx = (k_1, X, f, \mathcal{A}),\narx \models t_1\). Clearly there is \(k_1'\le k\), and the same can be obtained for \(k_2'\). Then \(t_                  1,t_2\) is said to be minimally mergeable when \(k_1'= k_2'\) and \(\exists \narx = (k_1', X, f, \mathcal{A}),\narx \models t_1, t_2\)
%     \end{itemize}
\end{definition}%
% \noindent
Determining whether two segments are minimally mergeable is analogous to that for the mergeable criterion, except that we need to enumerate from order $0$ to $k$ incrementally.

By applying the minimally mergeable criterion, $\traceDT_1$ and $\traceDT_2$ in \cref{example:aggressive-merging} can no longer be merged, nor can $\traceDT_3$ and $\traceDT_4$. However, we emphasize that, due to the limited information carried by the learning data, one can often not determine whether two traces should be merged or not (recall \cref{example:aggressive-merging}). Therefore, {\ourmethod} provides the two different 
merging criteria
% strategies (mergeable and minimally mergeable) 
as an option controlled by the users. In practice, we observe that the simple mergeable criterion \`{a} la \cref{def:mergeable} suffices to yield good clustering results.

Note that if the minimally mergeable criterion is chosen for segment 
clustering, then in mode characterization (\cref{subsec:mode-characterization}), we will instead find $\narxIns_i$ of order $\hat{k}_{S_i}$ that fits $S_i$. A similar argument on the correctness of mode characterization under the minimally mergeable criterion remains valid.

\section{Detailed Complexity Analysis of {\ourmethod}}\label{sec:appendix-complexity}

Below, we establish {\ourmethod}'s worst-case time complexity.

First, we explore the core procedure of NARX model fitting. Recall that the data matrices $\Dmatrix_i$ in the LLSQ optimization problem \cref{eqn:optimization} is of dimension $(\alpha + kn + m + 1) \times (\traceDTLen-k+1)$, where $\alpha$ and $k$ are respectively the number of nonlinear terms and the order of the template NARX model $\narx$, $n$ and $m$ are respectively the size of output variables $\varsO$ and input variables $\varsI$, and $(\traceDTLen+1)$ is the length of trace to be fitted. Let $d = (\alpha + kn + m + 1)$ be the row cardinality of $\Dmatrix_i$ (encoding the size of the target system).
%The complexity of evaluating the $\alpha$ nonlinear terms in the template $\narx$ is $\bigo{\alpha}$.
%Then, for a trace of length $\traceDTLen$, the complexity of constructing the matrix $\Dmatrix$ and $\Omatrix$ is $\bigO((\traceDTLen-k) \cdot (T_f+kn+m))$, and 
Then, the complexity of solving \cref{eqn:optimization} (e.g., using singular value decomposition \cite{DBLP:books/ox/07/GolubR07}) is $\bigO((\traceDTLen-k) \cdot d^2)$. Therefore, the total complexity of NARX fitting -- by solving $n$ optimization problems -- is $\bigO(n \cdot (\traceDTLen-k) \cdot d^2)$, denoted as $\bigO(T_N(\traceDTLen))$. This complexity reduces to $\bigO(T_N(\traceDTLen)) = \bigO(n \cdot \traceDTLen \cdot d^2)$ as $\traceDTLen \gg k$, which satisfies $\bigO(T_N(\traceDTLen_1)) + \bigO(T_N(\traceDTLen_2)) = \bigO(T_N(\traceDTLen_1 + \traceDTLen_2))$.

Then, we analyze the algorithmic complexity of each step in {\ourmethod}. Let $\abs{\data} = (\traceDTLen+1) \cdot M$ be the number of data points in the learning data. Then,
\begin{enumerate*}[label=(\Roman*)]
    \item \emph{Segmentation}: Each trace segment within the sliding window needs to be fitted, thus yielding the complexity $\bigO(\abs{\data} \cdot T_N(\windowSize))$, where $\windowSize$ is the window size;
    \item \emph{Clustering}: As we have to find mergeable segments\footnote{For the minimally mergeable criterion (see \cref{sec:appendix-mmergeable}), the complexity has to be multiplied by $k$ to take into account the enumeration.} by traversing each pair of them, the complexity is $\bigO(\sum_{\traceDT,\traceDT' \in \segments_{\data}} \! T_N(\abs{\traceDT} + \abs{\traceDT'})) = \bigO(T_N(\abs{\data})\cdot\abs{\segments_{\data}})$, and in the worst case,  $\bigO(T_N(\abs{\data})\cdot\abs{\data})$;
    \item \emph{Mode characterization}: To perform NARX fitting for each cluster of traces, the complexity is $\bigO(T_N(\abs{\data}))$;
    \item \emph{Guard learning}: The complexity of SVM is $\bigO((\abs{(q,q')^+} + \abs{(q,q')^-})^2 \cdot n)$ for each pair $(q,q')$ of modes, and the worst-case complexity of the whole SVM procedure is $\bigO(\abs{\data}^3 \cdot n)$;
    \item \emph{Reset learning}: %In general, the complexity of this step is negligible compared to the previous steps: 
    In the worst case, NARX model fitting needs to be performed on each segment with length $k+1$, and thus the complexity is $\bigO(\abs{\data} \cdot T_N(k))$.
\end{enumerate*}

As a consequence, the total \emph{worst-case} complexity of {\ourmethod} is $\bigO(\abs{\data}^3 \cdot n + T_N(\abs{\data})\cdot\abs{\data})$, i.e., \emph{polynomial} in the size of the learning data $\data$ and the size of the target system $d$. In practice, however, the complexity is much lower (bounded by $\bigO(T_N(\abs{\data})\cdot|\segments_\data| + \abs{\data} \cdot T_N(\windowSize))$).

\section{The Choice of NARX Templates}\label{sec:appendix-narx-templates}

The NARX templates in our experiments are constructed based on either (i) \emph{data characteristics}, e.g., including a high-order term such as $x^3$ by witnessing the gradually increasing period observed in \cref{fig:duffing-trace}; or (ii) \emph{physical intuition}, e.g., introducing the nonlinear term $\sin(x)$ while expecting a pendulum-like behavior for the \benchmark{oscillator} benchmark. In our experiments, the user-specified templates typically include a single nonlinear term (rational or trigonometric), even when the target dynamics involve multiple complex nonlinear terms.

\section{Extra Visualization of Case Studies}\label{sec:appendix-case-study}

\begin{figure}[H]
    \centering
    % \begin{subfigure}[b]{0.24\textwidth}
    %     \centering
    %     \includegraphics[width=\textwidth]{./fig/linear31.pdf}
    %     \caption{trace fidelity for \benchmark{complex\_tank}}
    %     \label{fig:complex_tank}
    % \end{subfigure}%
    % \hfill
    \begin{minipage}[b]{0.48\linewidth}
        \centering
        \includegraphics[width=\textwidth]{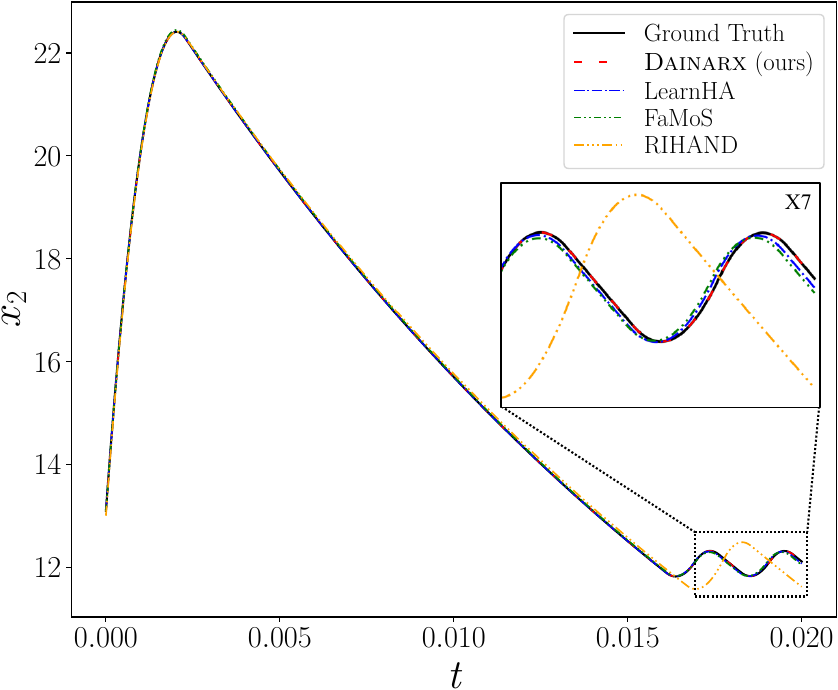}
        \caption{Traces of \benchmark{buck\_converter}.}
        \label{fig:buck_converter}
    \end{minipage}%
    \hfill
    \begin{minipage}[b]{0.44\linewidth}
        \centering
        \resizebox{\textwidth}{!}{%
            \begin{tikzpicture}[node distance=2cm, scale=1, transform shape]
                \node[draw, rounded corners, text width=3cm, align=center,minimum height=1.8cm] (q1) 
                {$q_1$\\[1mm]$x^{(1)} = 1$\\ $y^{(2)} = - 3 y^{(1)} -25y+27$};
                \node[draw, rounded corners, text width=3cm, align=center,minimum height=1.8cm, right=1.3cm of q1] (q2) 
                {$q_2$\\[1mm]$x^{(1)} = 1$\\ $y^{(2)} = - 3 y^{(1)} -25y +3$ };
                \node[draw, rounded corners, text width=3cm, align=center,minimum height=1.8cm, below=2.7cm of q2] (q3) 
                {$q_3$\\[1mm]$x^{(1)} = 1$\\ $y^{(2)} =3 y^{(1)} $};
                \node[draw, rounded corners, text width=3cm, align=center,minimum height=1.8cm, below=2.7cm of q1] (q4) 
                {$q_4$\\[1mm]$x^{(1)} = -4 $\\ $y^{(2)} = - 11 y^{(1)} -25y+20$};
                \draw[->, >=latex] (q1) to node[midway, above] {$x\geq5$} (q2);
                \draw[->, >=latex] (q2) to node[midway, right] {$x\geq10$} (q3);
                \draw[->, >=latex] (q3) to node[midway, above] {$x\geq12$} (q4);
                \draw[->, >=latex] (q4) to node[midway, left] {$x\leq0$} (q1);
            \end{tikzpicture}%
        }
        \vspace*{-0.7mm}
        \caption{HA of \benchmark{loop\_syst}.}
        \label{fig:loop}
    \end{minipage}%
    % \hfill
    % \begin{subfigure}[b]{0.26\textwidth}
    %     \centering
    %     \includegraphics[width=\textwidth]{./fig/case-annotated.pdf}
    %     % \begin{tikzpicture}[scale=1, transform shape]
    %     %     \draw (0, 0.) node[inner sep=0] {\includegraphics[width=\textwidth]{paper/fig/case.pdf}};
    %     %     \draw (-1.35, .79) node {\scriptsize \textcolor{customgreen}{$\traceDT_1$}};
    %     %     \draw (-0.36, -.3) node {\scriptsize \textcolor{customorange}{$\traceDT_2$}};
    %     %     \draw (.56, -.76) node {\scriptsize \textcolor{custompink}{$\traceDT_3$}};
    %     %     \draw (0.9, -.4) node {\scriptsize \textcolor{customcyan}{$\traceDT_4$}};
    %     %     \draw (1.8, -.12) node {\scriptsize \textcolor{customgreen}{$\traceDT_5$}};
    %     % \end{tikzpicture}
    %     \caption{segment. \& clustering of \benchmark{loop\_syst}}
    %     \label{fig:case_seg_cluster}
    % \end{subfigure}
    % \caption{Case studies demonstrating the effectiveness of {\ourmethod}.}
    % \label{fig:your_label}
\end{figure}

\section{Details on Execution Time}\label{sec:appendix-executiontime}

\cref{tab:detailed_time} summarizes the detailed execution time of different inference tools across all benchmarks. We record detailed comparisons in the major steps against LearnHA and FaMoS since they employ the same pipeline as {\ourmethod}.  We omit the explicit timings for the steps of mode characterization and reset learning as they are negligible compared to the other three steps.

\else
\fi

\end{document}